\documentclass[sigconf]{acmart}
\usepackage{graphicx}
\usepackage{xcolor}
\usepackage{multirow}
\usepackage{tabularx}
\usepackage{algorithm}
\usepackage{algorithmic}
\usepackage{subcaption}
\usepackage{amsthm}
\usepackage{enumitem}

\newcolumntype{C}{>{\centering\arraybackslash}X} % centered version of 'X' col. type
\newtheorem{problem}{Problem}

\newcommand{{\method}}{UGBA}

\copyrightyear{2023}
\acmYear{2023}
\setcopyright{acmlicensed}\acmConference[WWW '23]{Proceedings of the ACM
Web Conference 2023}{May 1--5, 2023}{Austin, TX, USA}
\acmBooktitle{Proceedings of the ACM Web Conference 2023 (WWW '23), May
1--5, 2023, Austin, TX, USA}
\acmPrice{15.00}
\acmDOI{10.1145/3543507.3583392}
\acmISBN{978-1-4503-9416-1/23/04}

\begin{document}
\fancyhead{}
\title{Unnoticeable Backdoor Attacks on Graph Neural Networks}

\author{Enyan Dai}
\email{emd5759@psu.edu}
\affiliation{
\institution{The Pennsylvania State University}
\city{State College}
\country{USA}}
\authornote{Both authors contribute equally to this paper.}

\author{Minhua Lin}
\email{mfl5681@psu.edu}
\affiliation{
\institution{The Pennsylvania State University}
\city{State College}
\country{USA}}
\authornotemark[1]

\author{Xiang Zhang}
\email{xzz89@psu.edu}
\affiliation{
\institution{The Pennsylvania State University}
\city{State College}
\country{USA}}

\author{Suhang Wang}
\email{szw494@psu.edu}
\affiliation{
    \institution{The Pennsylvania State University}
    \city{State College}
    \country{USA}
}



\begin{abstract}

Graph Neural Networks (GNNs) have achieved promising results in various tasks such as node classification and graph classification. Recent studies find that GNNs are vulnerable to adversarial attacks. However, effective backdoor attacks on graphs are still an open problem. In particular, backdoor attack poisons the graph by attaching triggers and the target class label to a set of nodes in the training graph. The backdoored GNNs trained on the poisoned graph will then be misled to predict test nodes to target class once attached with triggers. Though there are some initial efforts in graph backdoor attacks, our empirical analysis shows that they may require a large attack budget for effective backdoor attacks and the injected triggers can be easily detected and pruned. Therefore, in this paper, we study a novel problem of unnoticeable graph backdoor attacks with limited attack budget. To fully utilize the attack budget, we propose to deliberately select the nodes to inject triggers and target class labels in the poisoning phase. An adaptive trigger generator is deployed to obtain effective triggers that are difficult to be noticed. Extensive experiments on real-world datasets against various defense strategies demonstrate the effectiveness of our proposed method in conducting effective unnoticeable backdoor attacks.

% Graph Neural Networks (GNNs) have achieved promising results in various tasks such as node classification and graph classification. Recent studies find that GNNs are vulnerable to adversarial attacks. However, effective backdoor attacks on graphs are still an open problem. In particular, a backdoor attack poisons the graph by attaching triggers and the target class label to a set of nodes in the training graph. The backdoored GNNs trained on the poisoned graph will then be misled to predict test nodes to target class once attached with triggers. Though there are some initial efforts in graph backdoor attacks, our empirical analysis shows that they may require a large attack budget for effective backdoor attacks and the injected triggers can be easily detected and pruned. Therefore, in this paper, we study a novel problem of unnoticeable graph backdoor attacks with a limited attack budget. To fully utilize the attack budget, we propose to deliberately select the nodes to inject triggers and target class labels in the poisoning phase. An adaptive trigger generator is deployed to obtain effective triggers that are difficult to be noticed. Extensive experiments on real-world datasets against various defense strategies demonstrate the effectiveness of our proposed method in conducting effective unnoticeable backdoor attacks.

\end{abstract}

% \vskip -4em
\maketitle

\section{Introduction}
Graph-structured data are very pervasive in the real-world such as social networks~\cite{hamilton2017inductive}, finance system~\cite{wang2019semi}, and molecular graphs~\cite{irwin2012zinc}. Recently, graph neural networks (GNNs)~\cite{kipf2016semi,velivckovic2017graph} have shown promising results in modeling graphs. Generally, GNNs adopt the message-passing mechanism~\cite{kipf2016semi,xu2018powerful}, which updates a node's representation by aggregating information from its neighbors.  As a result, the node representations learned by GNNs can preserve node features, neighbors and local graph topology, which facilitate various tasks such as semi-supervised node classification and graph classification. 

Despite their great success of GNNs, GNNs are vulnerable to adversarial attacks and many graph attack methods have been investigated to fool the target GNN models to achieve adversarial goals~\cite{zugner2018adversarial,dai2018adversarial,xu2019topology}. Specifically, due to the utilization of graph structure with the message-passing mechanism, attackers can deliberately manipulate the graph structures and/or node features to mislead the GNNs for adversarial attacks. For instance, Nettack~\cite{zugner2018adversarial} iteratively modifies the connectivity of node pairs within attack budget to reduce the classification accuracy of GNNs on target nodes. 
However, the majority of existing attacks focus on graph manipulation attacks that require to calculate the optimal edges to be added/deleted for each target node. 
% \suhang{a graph that can achieve the following goal would be great: (1) show the issues of graph manipulation attack; (2) difference between graph manipulation and backdoor attack; and (3) issues of existing backdoor attacks} 
This will result in unaffordable time and space complexity on large-scale datasets, i.e, $O(N^2)$ where $N$ is the number of nodes~\cite{zugner2018adversarial,dai2018adversarial,dai2022comprehensive}. 
In addition, manipulating links and attributes of existing nodes is impractical as these nodes/individuals are not controlled by the attacker~\cite{sun2020nipa}.%\suhang{@Enyan, graph attack can not address this as graph attack need to (i) connect trigger to the target nodes; and (ii) assign target class to the target nodes during training. Need to rephrase this.}. 

\begin{figure}[t]
\centering
\includegraphics[width=0.93\linewidth]{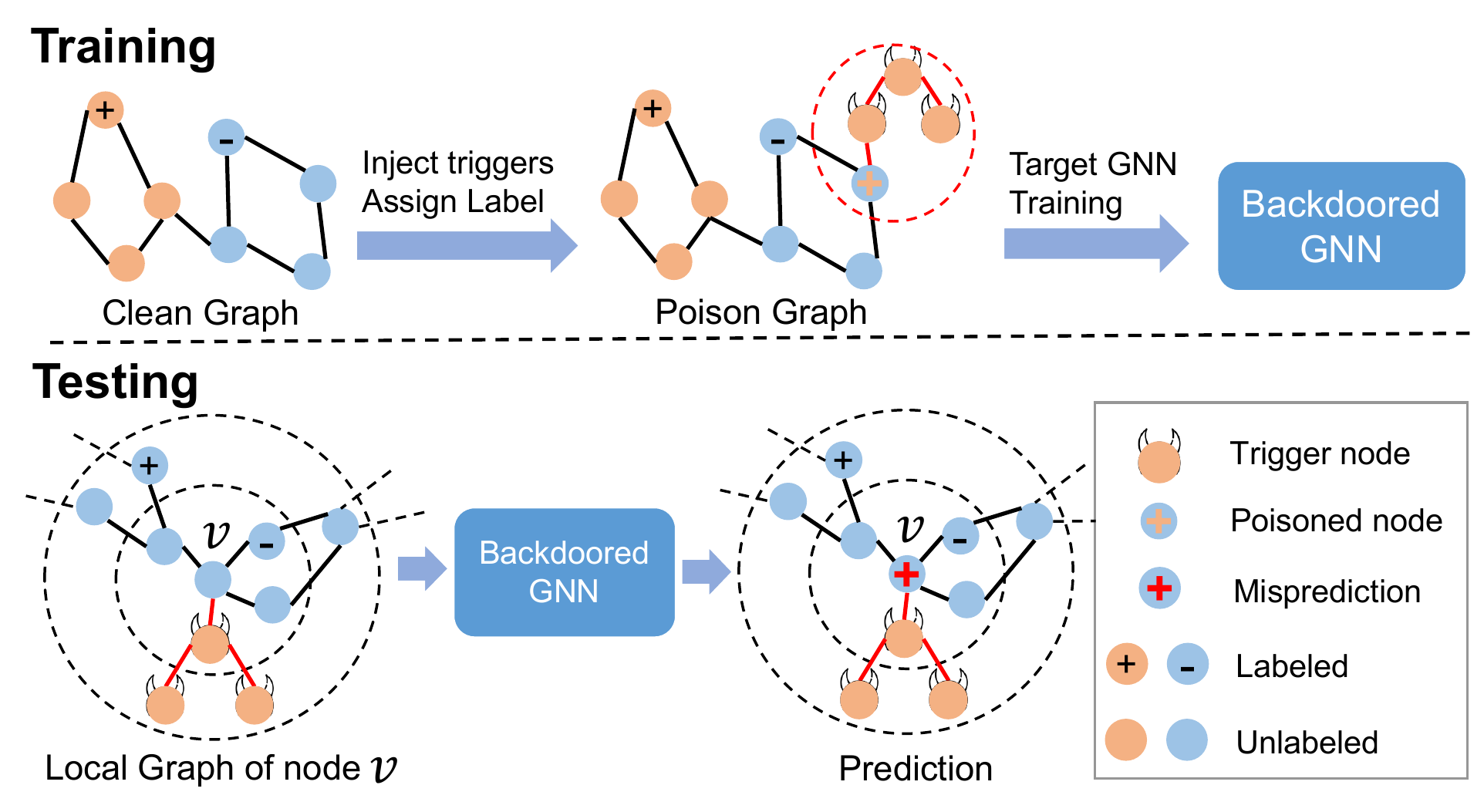}
% \vspace{-3mm}
\vspace{-1.2em}
\caption{General framework of graph backdoor attack.} 
\label{fig:general_framework}
\vskip -1.5em
\end{figure}

To address the aforementioned issues, one promising direction is to develop backdoor attacks on graphs. Fig.~\ref{fig:general_framework} gives an illustration of backdoor attack on graphs, where a small set of nodes denoted as poisoned samples will be attached with triggers and assigned the label of target class.  The model trained on the poisoned graph will link the trigger with the target class. As a result, the target nodes will be predicted as the target class once they are attached with the triggers during the inference phase. The trigger can be either predefined or obtained from trigger generator. 
\textit{Firstly}, the computation cost of backdoor attacks is limited compared with graph manipulation attacks, which paves us a way for an efficient target attack on large-scale graphs. 
For predefined triggers, nearly no computation cost is required. When a trigger generator is adopted, optimizing the trigger generator only needs the gradients from the poisoned samples. 
\textit{Secondly}, once the backdoor is injected to the target GNNs, the predictions on new target nodes can be easily controlled by attaching generated triggers instead of an additional optimization process as graph manipulation attacks. This will especially benefit the targeted attack on inductive node classification, which widely exists in real-world scenarios. For example, TikTok graph will often incorporate new users and predict labels of them with a trained model. \textit{Thirdly}, compared with revising the links between existing users, it is relatively easy to inject triggers and malicious labels in backdoor attacks. Take malicious user detection on social networks as an example, many labels are collected from reports of users. In this case, malicious labels could be easily assigned by attackers. As for the trigger attachment, it can be achieved by linking a set of fake accounts to the users.

Recently, \citet{zhang2021backdoor} firstly investigate a graph backdoor attack that uses randomly generated graphs as triggers. A trigger generator is adopted in~\cite{xi2021graph} to get more powerful sample-specific triggers. However, these methods have unnoticeablity issues in the following two aspects. 
\textit{Firstly}, our empirical analysis in Sec.~\ref{sec:3.3.1} shows that existing methods need a large budget to conduct effective backdoor attacks on large-scale graphs, i.e., they need to attach the backdoor triggers to a large number of nodes in the training graph so that a model trained on the graph will be fooled to assign target label to nodes attached with the backdoor trigger. This largely increases the risk of being detected. 
\textit{Secondly}, the generated triggers of these methods can be easily identified and destroyed. Specifically, real-world graphs such as social networks generally follow homophily assumption, i.e., similar nodes are more likely to be connected; while in existing graph backdoor attacks, the edges linking triggers and poisoned nodes and edges inside the triggers are not guaranteed with the property of connecting nodes with high similarity scores. Thus, the triggers and assigned malicious labels can be eliminated by pruning edges linking dissimilar nodes and discarding labels of involved nodes, which is verified in Sec~\ref{sec:3.3.2}. Thus, developing an effective unnoticeable graph backdoor attack with limited attack budget is important. However, graph backdoor attack is still in its early stage and there is no existing work on unnoticeable graph backdoor attack with limited attack budget.

Therefore, in this paper, we study a novel and important problem of developing an effective unnoticeable graph backdoor attack with limited attack budget in terms of the number of poisoned nodes. In essence, we are faced with two challenges: (i) how to fully utilize the limited budget in poisoned samples for graph backdoor attacks; (ii) how to obtain triggers that are powerful and difficult to be detected. In an attempt to address these challenges, we proposed a novel framework \underline{U}nnoticeable \underline{G}raph \underline{B}ackdoor \underline{A}ttack (UGBA)\footnote{https://github.com/ventr1c/UGBA}. To better utilize the attack budget, {\method} proposes to attach triggers with crucial representative nodes with a novel poisoned node selection algorithm. And an adaptive trigger generator is deployed in {\method} to obtain powerful unnoticeable trigger that exhibits high similarity with each target node and maintains high attack success rate. In summary, our main contributions are:
\begin{itemize}[leftmargin=*]
    \item We study a novel problem of promoting unnoticeablity of graph backdoor attacks in generated triggers and attack budget;
    \item We empirically verify that a simple strategy of edge pruning and label discarding can largely degrade existing backdoor attacks;
    \item We design a framework {\method} that deliberately selects poisoned samples and learn effective unnoticeable triggers to achieve unnoticeable graph backdoor attack under limited budget; and
    \item Extensive experiments on large-scale graph datasets demonstrate the effectiveness of our proposed method in unnoticeably backdooring different GNN models with limited attack budget. 
\end{itemize}

\section{Related Works}
% \enyan{Minhua please fill in these section}
\subsection{Graph Neural Networks}
Graph Neural Networks (GNNs)~\cite{kipf2016semi,velivckovic2017graph,ying2018graph,bongini2021molecular} have shown remarkable ability in modeling graph-structured data, which benefits various applications such as recommendation system~\cite{ying2018graph}, drug discovery~\cite{bongini2021molecular} and traffic analysis~\cite{zhao2020semi}. Generally, the success of GNNs relies on the message-passing strategy, which updates a node's representation by recursively aggregating and combining features from neighboring nodes. For instance, in each layer of GCN~\cite{kipf2016semi} the representations of neighbors and the center node will be averaged followed by a non-linear transformation such as ReLU. 
% GAT~\cite{velivckovic2017graph} considers an attention mechanism to determine the weights of each node's neighbors, which can better aggregate the neighborhood information.
Recently, many GNN models are proposed to further improve the performance of GNNs~\cite{chen2020simple,kim2021find,zhu2020self,zeng2020graphsaint,dai2021nrgnn,dai2022towards,dai2022comprehensive}. For example, 
% deep GNNs~\cite{li2019deepgcns,chen2020simple,li2021gnn1000} are proposed to alleviate the oversmoothing issue of GCN and incorporate more hops of neighbors. Moreover, 
self-supervised GNNs~\cite{kim2021find,zhu2020self,qiu2020gcc} are investigated to reduce the need of labeled nodes. Works that improve  fairness~\cite{dai2021say}, robustness~\cite{dai2021nrgnn,dai2022towards} and explainability~\cite{dai2021towards,ying2019gnnexplainer,zhang2022protgnn} of GNNs are explored. And GNN models for heterophilic graphs are also desgined~\cite{xu2022hp,dai2022labelwise}.

\subsection{Attacks on Graph Neural Networks}
% Existing studies showed that GNNs are vulnerable to adversarial attacks that perturb the graph structures and node attributes.
According to the stages the attack occurs, adversarial attacks on GNNs can be divided into poisoning attack~\cite{zugner2018adversarial, zugner2019adversarial, sun2020nipa} and evasion attack~\cite{dai2018adversarial,xu2019topology, wang2020evasion,tao2021single,bojchevski2019adversarial,chang2020restricted}. In poisoning attacks, the attackers aim to perturb the training graph before GNNs are trained such that a GNN model trained on the poisoned dataset will have a low prediction accuracy on test samples. For example, Nettack~\cite{zugner2018adversarial} employs a tractable surrogate model to conduct a targeted poisoning attack by learning perturbation against the surrogate model. 
% Bojchevski et al.~\cite{bojchevski2019adversarial} exploit the eigenvalue perturbation theory to efficiently poison the graph structure. 
% Metatack~\cite{zugner2019adversarial} introduces meta-learning to solve a bi-level problem on the non-targeted attack, which treats the graph structure matrix as a hyperparameter to optimize. 
Evasion attacks add perturbation in the test stage, where the GNN model has been well trained and cannot be modified by attackers. Optimizing the perturbation of graph structures by gradient descent~\cite{xu2019topology} and reinforcement learnings~\cite{dai2018adversarial, ma2021rewiring} have been explored. Evasion attacks through graph injection~\cite{zou2021tdgia,koh2017understanding} are also investigated.

Backdoor attacks are still rarely explored on GNNs~\cite{zhang2021backdoor,xi2021graph}. Backdoor attacks generally attach backdoor triggers to the training data and assign the target label to samples with trigger. Then a model trained on the poisoned data will be misled if backdoors are activated by the trigger-embedded test samples. \citeauthor{zhang2021backdoor}~\cite{zhang2021backdoor} propose a subgraph-based backdoor attack on GNNs by injecting randomly generated universal triggers to some training samples. \citeauthor{xi2021graph}~\cite{xi2021graph} adopt a trigger generator to learn to generate adaptive trigger for different samples. \citeauthor{sheng2021backdoor}~\cite{sheng2021backdoor} propose to select the nodes with high degree and closeness centrality. ~\citeauthor{xu2022poster}~\cite{xu2022poster} improve the unnoticeability by assigning triggers without change labels of poisoned samples. Our proposed method is inherently different from these methods as (i) we can generate unnoticeable adaptive triggers to simultaneously maintain the effectiveness and bypass the potential trigger detection defense based on feature similarity of linked nodes; (ii) we design a novel clustering-based node selection algorithm to further reduce the required attack budget.
\section{Preliminary Analysis}
In this section, we present preliminaries of backdoor attacks on graphs and show unnoticeablity issues of existing backdoor attacks.

\subsection{Notations} \label{sec:notations}
We use $\mathcal{G}=(\mathcal{V},\mathcal{E}, \mathbf{X})$ to denote an attributed graph, where $\mathcal{V}=\{v_1,\dots,v_N\}$ is the set of $N$ nodes, $\mathcal{E} \subseteq \mathcal{V} \times \mathcal{V}$ is the set of edges, and $\mathbf{X}=\{\mathbf{x}_1,...,\mathbf{x}_N\}$ is the set of node attributes with $\mathbf{x}_i$ being the node attribute of $v_i$. $\mathbf{A} \in \mathbb{R}^{N \times N}$ is the adjacency matrix of the graph $\mathcal{G}$, where $\mathbf{A}_{ij}=1$ if nodes ${v}_i$ and ${v}_j$ are connected; otherwise $\mathbf{A}_{ij}=0$. In this paper, we focus on a semi-supervised node classification task in the inductive setting, which widely exists in real-world applications. For instance, GNNs trained on social networks often need to conduct predictions on newly enrolled users to provide service. Specifically, in inductive node classification, a small set of nodes $\mathcal{V}_L \subseteq \mathcal{V}$ in the training graph are provided with labels $\mathcal{Y}_L=\{y_1,\dots,y_{N_L}\}$. The test nodes $\mathcal{V}_T$ are not covered in the training graph $\mathcal{G}$, i.e., $\mathcal{V}_T \cap \mathcal{V} = \emptyset$. 

\subsection{Preliminaries of Graph Backdoor Attacks}
\subsubsection{Threat Model} In this section, we introduce the threat model.

\noindent{\bf{Attacker's Goal}}: The goal of the adversary is to mislead the GNN model to classify target nodes attached with the triggers as target class. Simultaneously, the attacked GNN model should behave normally for clean nodes without triggers attached.

\noindent{\bf{Attacker's Knowledge and Capability}}: As the setting of most poisoned attacks, the training data of the target model is available for attackers. The information of the target GNN models including model architecture is unknown to the attacker. Attackers are capable of attaching triggers and labels to nodes within a budget before the training of target models to poison graphs. During the inference phase, attackers can attach triggers to the target test node. 

\subsubsection{General Framework of Graph Backdoor Attacks}
The key idea of the backdoor attacks is to associate the trigger with the target class in the training data to mislead target models. 
As Fig.~\ref{fig:general_framework} shows, during the poisoning phase, the attacker will attach a trigger $g$ to a set of poisoned nodes $\mathcal{V}_P \subseteq \mathcal{V}$ and assign $\mathcal{V}_P$ with target class label $y_t$, resulting a backdoored dataset. Generally, the poisoned node set $\mathcal{V}_P$ is randomly selected.
The GNNs trained on the backdoored dataset will be optimized to predict the poisoned nodes $\mathcal{V}_P$ attached with the trigger $g$ as target class $y_t$, which will force the target GNN to correlate the existence of the trigger $g$ in neighbors with the target class. In the test phase, the attacker can attach the trigger $g$ to a test node $v$ to make $v$ classified as the target class by backdoored GNN. Some initial efforts~\cite{zhang2021backdoor,xi2021graph} have been made for graph backdoor attacks. Specifically, SBA~\cite{zhang2021backdoor} directly injects designed sub-graphs as triggers. And GTA~\cite{xi2021graph} adopts a trigger generator to learn optimal sample-specific triggers.
% \enyan{SBA uses human designed trigger. GTA uses optimized sample-specific triggers under bi-level optimization. Should we use bi-level optimization objective function to introduce or a unified framework figure as doing now?} \suhang{a general figure for unified framework is enough. Followed by that, you can briefly mention SBA and GTA}.

\subsection{Unnoticeability of Graph Backdoor Attacks}
% Several works~\cite{xi2021graph,zhang2021backdoor} demonstrate the vulnerability of GNNs to backdoor attacks. However, the unnoticeability of them has not been explored.
In this subsection, we analyze the unnoticeability of existing graph backdoor attacks in terms of the required number of poisoned samples and the difficulty of trigger detection.

\subsubsection{Size of Poisoned Nodes} \label{sec:3.3.1}
In backdoor attacks, a set of poisoned nodes $\mathcal{V}_P$ will be attached triggers and target class labels to conduct attacks. However, as large-scale graphs can provide abundant information for training GNNs, the attacker may need to inject a large number of triggers and malicious labels to mislead the target GNN to correlate the trigger with target class, which puts backdoor attack at the risk of being noticed. 
To verify this, we analyze how the size of poisoned nodes affects the attack success rate of the state-of-the-art graph backdoor attacks, i.e., SBA-Gene, SBA-Samp~\cite{zhang2021backdoor}, and GTA~\cite{xi2021graph} on a large node classification dataset, i.e., OGB-arxiv~\cite{hu2020ogb}. Detailed descriptions of these methods can be found in Sec.~\ref{sec:baseline}. We vary $|\mathcal{V}_P|$ as $\{80,240,800,2400\}$. The size of trigger is limited to contain three nodes. The architecture of target model is GraphSage~\cite{hamilton2017inductive}. The attack success rate (ASR) results are presented in Tabble~\ref{tab:pre_poison}. From the table, we can observe that all methods especially SBA-Gen and SBA-Samp achieve poor attack results with limited budget such as 80 and 240 in $\mathcal{V}_P$. This is because (i) SBA-Gen and SBA-Samp utilize handcrafted triggers which is not effective; (ii) Though GTA uses learned sample-specific trigger, similar to SBA-Gen and SBA-Samp, the selection of poisoned nodes is random and the budget is not well utilized. 
% require a large $|\mathcal{V}_P|$ to conduct effective backdoor attacks; (ii) due to the deployment of sample-specific trigger optimization, GTA performs better than the others but still give poor attack performance when the size of $\mathcal{V}_P$ is limited to be 80. With a large size of poisoned nodes, the cost and risk of being caught will be high. 
Thus, it is necessary to develop graph backdoor attack methods that can generate effective triggers and fully exploit the attack budget.

\begin{table}[t]
    \centering
    \small
    \caption{Impacts of $|\mathcal{V}_P|$ to ASR (\%) of backdoor attacks.}
    \vskip -1.2em
    \begin{tabularx}{0.95\linewidth}{p{0.16\linewidth}CCCCC}
        \toprule
         $|\mathcal{V}_P|$ & 80 & 240 & 400 & 800 & 2400 \\
         \midrule
         SBA-Samp & 0.06  & 1.7 & 10.8 & 34.5 & 75.5 \\
         SBA-Gen  &  0.08 & 18.1 & 32.1 & 54.3 & 85.9 \\
         GTA      & 37.4  & 62.4 & 72.4 & 82.7 & 94.8\\
         \bottomrule
    \end{tabularx}
    \vskip -1em
    \label{tab:pre_poison}
\end{table}

\begin{table}[t]
    \centering
    \small
    \vskip -0.8em
    \caption{Results of backdoor defense (Attack Success Rate (\%) | Clean Accuracy (\%)) on Ogb-arxiv dataset.}
    \vskip -1.5em
    \begin{tabularx}{0.98\linewidth}{p{0.12\linewidth}CCCC}
        \toprule
        Defense & Clean &SBA-Samp & SBA-Gen & GTA \\
        \midrule
        None     & 65.5 & 61.0 | 65.1 & 70.8 | 65.2 & 94.8 | 65.6 \\
        Prune    & 62.2 & \hspace{2pt} 8.9 | 64.0 & 31.2 | 64.0  & \hspace{2pt} 1.4 | 64.5\\
        Prune+LD & 62.6 & \hspace{2pt} 3.2 | 64.0 & 15.3 | 63.8  & 0.04 | 64.1 \\
         \bottomrule
    \end{tabularx}
    \vskip -1em
    \label{tab:pre_prune}
\end{table}

\subsubsection{Detection of Triggers} \label{sec:3.3.2}
Real-world graphs such as social networks generally show homophily property, i.e, nodes with similar attributes are connected by edges. For existing backdoor attacks, the attributes of triggers may differ a lot from the attached poisoned nodes. The connections within trigger may also violate homophily property. Therefore,  the negative effects of injected triggers and target labels might be reduced by eliminating edges linking dissimilar nodes and labels of involved nodes. To verify this, we evaluate two strategies to defend  against backdoor attacks:
\begin{itemize}[leftmargin=*]
    \item  \textbf{Prune}: We prune edges linking nodes with low cosine similarity. As edges created by the backdoor attacker may link dissimilar nodes, the trigger structure and attachment edge can be destroyed.
    \item \textbf{Prune+LD}: To reduce the influence of dirty labels of poisoned nodes, besides pruning, we also discard the labels of the nodes linked by dissimilar edges.
\end{itemize}
Experimental results on Ogb-arxiv with  $|\mathcal{V}_P|$ set as 2400 are presented in Table~\ref{tab:pre_prune}. Other settings are the same as Sec.~\ref{sec:3.3.1}. For Prune and Prune+LD, the threshold is to filter out edges with lowest 10\% cosine similarity scores. More results on other datasets can be found in Table~\ref{tab:RQ1_table}. The accuracy of the backdoored GNN on clean test set is also reported in Table~\ref{tab:pre_prune} to show how the defense strategies affect the prediction performance. Accuracy on a clean graph without any attacks is reported as reference. All the results are average scores of 5 runs. We can observe from Tab.~\ref{tab:pre_prune} that (i) ASR drops dramatically with the proposed two strategies of prune and prune+LD; (ii) the impact of the proposed strategies on prediction accuracy is negligible. This demonstrates that the used triggers by existing backdoor attacks can be easily mitigated.

\section{Problem Formulation}
% \enyan{Not sure whether }
Our preliminary analysis verifies that existing backdoor attacks (i) require a large attack budget on large datasets; and (ii) the injected triggers can be easily detected. To alleviate these two issues, we propose to investigate a novel unnoticeable graph backdoor attack problem that can unnoticeablely backdoor various target GNNs with limited attack budget. Specifically, we enhance the general graph backdoor attack model from the following two aspects.
\vskip 0.2em
\noindent \textbf{Selection of Poisoned Nodes $\mathcal{V}_P$}: In the attack model of current graph backdoor attacks, the poisoned node set $\mathcal{V}_P$ is randomly selected. However, in this way, it is likely the budget is wasted in some useless poisoned nodes. For example, the attacker may repeatedly poison nodes from the same cluster that have very similar pattern, which is unnecessary. Alternatively, to fully utilize the attack budget, we will deliberately select the most useful poisoned nodes $\mathcal{V}_P \subseteq \mathcal{V}$ in unnoticeable backdoor attack. 
\vskip 0.2em
\noindent \textbf{Unnoticeable Constraint on Triggers}: 
As the preliminary analysis shows, dissimilarity among trigger nodes and poisoned nodes makes the attack easy to be detected. Hence, it is necessary to obtain adaptive triggers that are similar to the poisoned nodes or target nodes. 
In addition, edges within triggers should also be enforced to link similar nodes to avoid being damaged by  pruning strategy. Such adaptive trigger can be given by an adaptive generator. 
Let $\mathcal{E}_B^i$ denote the edge set that contain edges inside trigger $g_i$ and edge attaching trigger $g_i$ and node $v_i$. The unnoticeable constraint on the generated adaptive triggers can be formally written as: 
\begin{equation}
    \min_{(u,v) \in \mathcal{E}_B^i} sim(u,v) \geq T, 
    \label{eq:unc}
\end{equation}
where $sim$ denotes the cosine similarity between node features and $T$ is a relatively high threshold of the cosine similarity which can be tuned based on datasets. 

In node classification with GNNs, the prediction is given based on the computation graph of the node. Thus, the clean prediction on node $v_i$ can be written as $f_\theta(\mathcal{G}^i_C)$, where $\mathcal{G}^i_C$ denotes the clean computation graph of node $v_i$. For a node $v_i$ attached with the adaptive trigger $g_i$, the predictive label will be given by $f_\theta(a(\mathcal{G}^i_C, g_i))$, where $a(\cdot)$ denotes the operation of trigger attachment.  Then, with the above descriptions and notations in Sec~\ref{sec:notations}.
we can formulate the unnoticeable graph backdoor attack by:
\begin{problem} Given a clean attributed graph $\mathcal{G}=(\mathcal{V},\mathcal{E}, \mathbf{X})$ with a set of nodes $\mathcal{V}_L$ provided with labels $\mathcal{Y}_L$, we aim to learn an adaptive trigger generator $f_g: v_i \rightarrow g_i$ and effectively select a set of nodes $\mathcal{V}_P$ within budget to attach triggers and labels so that a GNN $f$ trained on the poisoned graph will classify the test node attached with the trigger to the target class $y_t$ by solving: 
\begin{equation}
\begin{aligned}
    \min_{\mathcal{V}_P, \theta_{g}} & \sum_{v_i \in \mathcal{V}_U} l(f_{\theta^*}(a(\mathcal{G}_C^i,g_i)), y_t) \\
    s.t. ~~  & \theta^* = \mathop{\arg\min}_{\theta}  \sum_{v_i \in \mathcal{V}_L}l(f_{\theta}(\mathcal{G}_C^i), y_i) + \sum_{v_i \in \mathcal{V}_P}l(f_{\theta}(a(\mathcal{G}_C^i,g_i)), y_t), \\
    & \forall v_i \in \mathcal{V}_P \cup \mathcal{V}_{T}, ~~\text{$g_i$ meets Eq.(\ref{eq:unc}) and }|g_i| < \Delta_{g} \\
    & |\mathcal{V}_P| \leq \Delta_{P} 
    \label{eq:opt}
\end{aligned}
\end{equation}
where $l(\cdot)$ represents the cross entropy loss and $\theta_g$ denotes the parameters of the adaptive trigger generator $f_g$. In the constraints, the node size of trigger $|g_i|$ is limited by $\Delta_{g}$, and the size of poisoned nodes is limited by $\Delta_{P}$. 
The architecture of the target GNN $f$ is unavailable and may adapt various defense methods.
\end{problem}
In transductive setting, $\mathcal{V}_U$ would be the target nodes. However, we focus on inductive setting where $\mathcal{V}_T$ is not available for the optimization. Hence, $\mathcal{V}_U$ would be $\mathcal{V} \backslash \mathcal{V}_L$ to ensure the attacks can be effective for various types of target nodes.
% \enyan{feel necessary to define our graph backdoor attack with node selection and unnoticeable constraint here. This is also a general form of attack paper: formulate the optimization problem and give the framework. But I am concerned about whether the concept of poisoned selection and unnoticeable constraint on triggers is unclear here}
\section{Methodology}
\begin{figure}[t]
% \vspace{-2mm}
\centering
% \vspace{-4mm}
\vskip -1em
\includegraphics[width=0.95\linewidth]{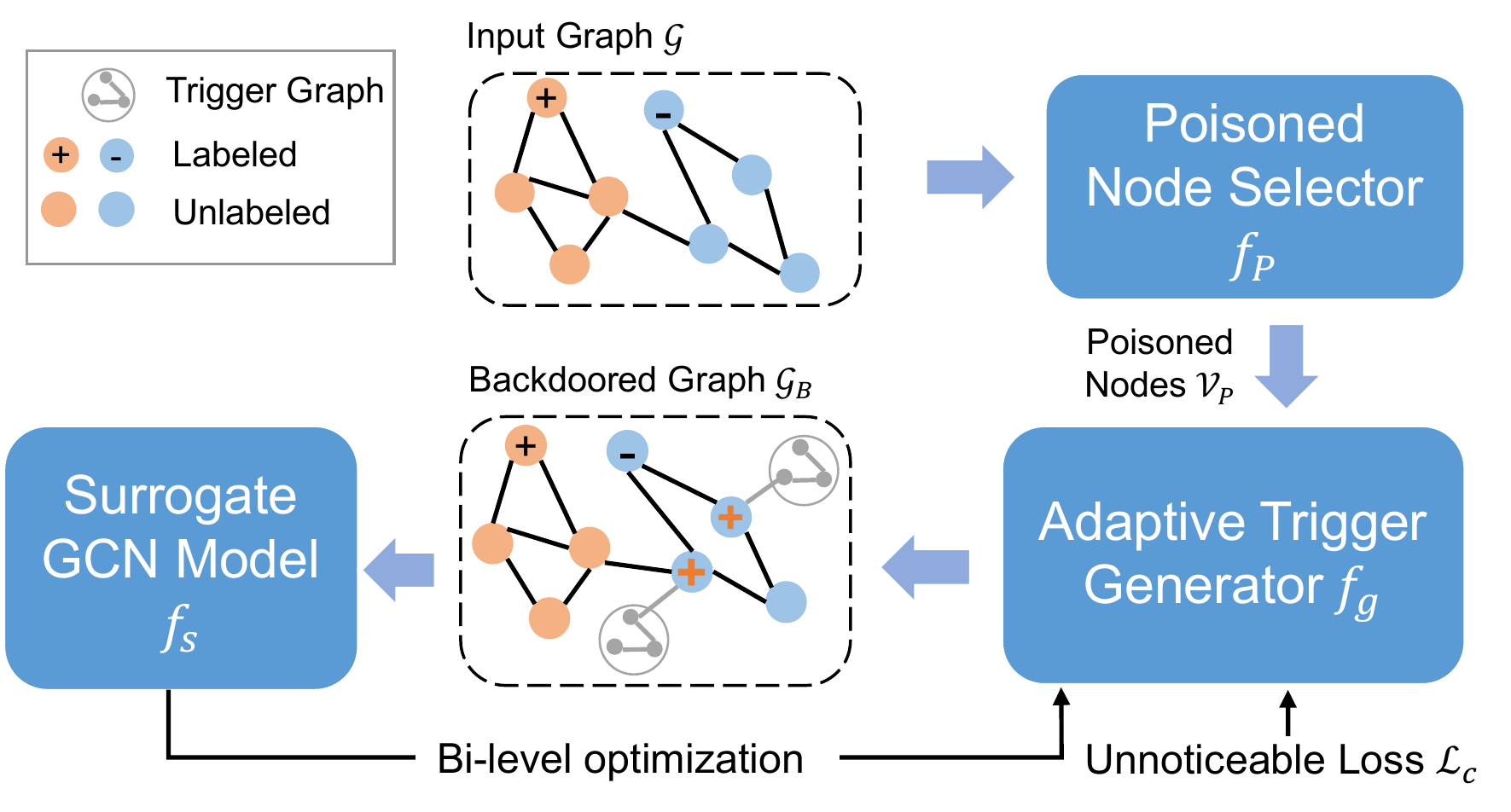}
\vspace{-1.5em}
\caption{An overview of proposed UGBA.} 
\label{fig:overall_framework}
%\vspace{+2mm}
\vspace{-1.5em}
\end{figure}

In this section, we present the details of {\method} which aims to optimize Eq.(\ref{eq:opt}) to conduct effective and unnoticeable graph backdoor attacks.  
Since it is challenging and computationally expensive to jointly optimize the selection of poisoned nodes $\mathcal{V}_P$ and the trigger generator, {\method} splits the optimization process into two steps: poisoned node selection and adaptive trigger generator learning. 
Two challenges remain to be addressed: (i) how to select the poisoned nodes that are most useful for backdoor attacks; (ii) how to learn the adaptive trigger generator to obtain triggers that meet unnoticeable constraint and maintain a high success rate in backdoor attack; To address these challenges, a novel framework of {\method} is proposed, which is illustrated in Fig.~\ref{fig:overall_framework}. {\method} is composed of a poisoned node selector $f_P$, an adaptive trigger generator $f_g$, and a surrogate GCN model $f_s$. Specifically, the poisoned node selector takes the graph $\mathcal{G}$ as input and applies a novel metric to select nodes with representative patterns in features and local structures as poisoned nodes. An adaptive trigger generator $f_g$ is applied with a differentiable unnoticeable constraint to give unnoticeable triggers for selected poisoned nodes $\mathcal{V}_P$ to fool $f_s$. To guarantee the effectiveness of the generated adaptive triggers on various test nodes, a bi-level optimization with a surrogate GCN model is applied.

\subsection{Poisoned Node Selection}
In this subsection, we give the details of the node selection algorithm. Intuitively, if nodes with representative features and local structures are predicted to the target class $y_t$ after being attached with triggers, other nodes are also very likely to be conducted successful backdoor attacks. Therefore, we propose to select diverse and representative nodes in the graph as poisoned nodes, which enforce the target GNN to predict the representative nodes attached with triggers to be target class $y_t$.

One straightforward way to obtain the representative nodes is to conduct clustering on the node features. However, it fails to consider the graph topology which is crucial for graph-structured data. Therefore, we propose to train a GCN encoder with the node labels to obtain representations that capture both attribute and  structure information. Then, for each class, we can select representative nodes using the clustering algorithm on learned representations. Specifically, the node representations and labels can be obtained as:
\begin{equation}
    \mathbf{H} = GCN(\mathbf{A}, \mathbf{X}), \quad \hat{\mathbf{Y}} = \text{softmax}(\mathbf{W} \cdot \mathbf{H}),
\end{equation}
where $\mathbf{W}$ denotes the learnable weight matrix for classification. 
The training process of the GCN encoder can be written as:
\begin{equation}
    \min_{\theta_E, \mathbf{W}} \sum_{v_i \in \mathcal{V}_L} l(\hat{y}_i, y_i)
    \label{eq:train_select}
\end{equation}
where $\theta_E$ denotes the parameters of GCN encoder, $l(\cdot)$ is the cross entropy loss, and $y_i$ is the label of node $v_i$. $\hat{y}_i$ is the prediction of $v_i$.

With the GCN encoder trained in Eq.~(\ref{eq:train_select}), we can obtain the node representations and conduct clustering to obtain the representative nodes for each class. Here, to guarantee the diversity of the obtained representative nodes, we separately apply K-Means to cluster $\{\mathbf{h}_i: \hat{y}_i=l\}$ on each class $l$ other than the target class $y_t$, where $\mathbf{h}_i$ denote the representation of node $v_i \in \mathcal{V}/\mathcal{V}_L$.
Nodes nearer to the centroid of each cluster are more representative. However, the node nearest to the centroid may have a high degree. Injecting the malicious label to high-degree nodes may lead to a significant decrease in prediction performance as the negative effect will be propagated to its neighbors, which may make the attack noticeable. Hence,  we propose  a metric that balances the representativeness and negative effects on the prediction performance. Let $\mathbf{h}_c^k$ denote the center of the $k$-th cluster. Then for a node $v_i^k$ belonging to the $k$-th cluster, the metric score can be computed by:
\begin{equation}
    m(v_i) = ||\mathbf{h}_i^k - \mathbf{h}_c^k||_2 + \lambda \cdot deg(v_i^k)
    \label{eq:node_sel_score}
\end{equation}
where $\lambda$ is to control the contribution of the degree in node selection. After getting each node's score, we select nodes with top-$n$ highest scores in each cluster to satisfy the budget, where $n=\frac{\Delta_P}{(C-1)K}$. 
% Such informative and representative nodes 

\subsection{Adaptive Trigger Generator}
Once the poisoned node set $\mathcal{V}_P$ is determined, the next step is to generate adaptive triggers with $f_g$ to poison the dataset. To guarantee the unnoticeability of the generated triggers, we propose a differentiable unnoticeable loss. We apply a bi-level optimization between the adaptive generator $f_g$ and the surrogate model $f_s$ to ensure high success rate on various test samples. Next, we give the details of trigger generator $f_g$, differentiable unnoticeable 
loss, and the bi-level optimization with $f_s$.

\vspace{0.2em}
\noindent \textbf{Design of Adaptive Trigger Generator.}  To generate adaptive triggers that are similar to the attached nodes, the adaptive trigger generator $f_g$ takes the node features of the target node as input. Specifically, we adopt an MLP to simultaneously generate node features and structure of the trigger for node $v_i$ by:
\begin{equation}
    \mathbf{h}_i^m = \text{MLP}(\mathbf{x}_i), \quad \mathbf{X}_i^g = \mathbf{W}_f \cdot \mathbf{h}_i^m, \quad \mathbf{A}_i^g = \mathbf{W}_a \cdot \mathbf{h}_i^m,
\end{equation}
where $\mathbf{x}_i$ is the node features of $v_i$. $\mathbf{W}_f$ and $\mathbf{W}_a$ are learnable parameters for feature and structure generation, respectively. $\mathbf{X}_i^g \in \mathbb{R}^{s \times d}$ is the synthetic features of the trigger nodes, where $s$ and $d$ represent the size of the generated trigger and the dimension of features, respectively.  $\mathbf{A}_i^g \in \mathbb{R}^{s \times s}$ is the adjacency matrix of the generated trigger. As the real-world graph is generally discrete, following the binary neural network~\cite{hubara2016binarized}, we binarize the continuous adjacency matrix  $\mathbf{A}_i^g$ in the forward computation; while the continuous value is used in backward propagation. With the generated trigger $g_i=(\mathbf{X}_i^g, \mathbf{A}_i^g)$, we link it to node $v_i \in \mathcal{V}_P$ and assign target class label $y_t$ to build backdoored dataset. In the inference phase, the trigger generated by $f_g$ will be attached to the test node $v_i \in \mathcal{V}_T$ to lead backdoored GNN to  predict it as target class $y_t$.

\vspace{0.2em}
\noindent \textbf{Differentiable Unnoticeable Loss.} The adaptive trigger generator $f_g$ aims to produce the triggers that meet the Eq.(\ref{eq:unc}) for unnoticeable trigger injection. The key idea is to ensure the poisoned node or test node $v_i$ is connected to a trigger node with high cosine similarity to avoid trigger elimination. And within the generated trigger $g_i$, the connected trigger nodes should also exhibit high similarity. Thus, we design a differentiable unnoticeable loss to help optimize the adaptive trigger generator $f_g$. Let $\mathcal{E}_B^i$ denote the edge set that contains edges inside trigger $g_i$ and edge attaching trigger $g_i$ and node $v_i$, the unnoticeable loss can be written as:
\begin{equation}
    \min_{\theta_g} \mathcal{L}_c = \sum_{v_i \in \mathcal{V}} \sum_{(v_j,v_k) \in \mathcal{E}_B^i} \max(0,T - sim(v_j,v_k)),
    \label{eq:l_c}
\end{equation}
where $T$ denotes the threshold of the similarity, and $\theta_g$ represents the parameters of $f_g$. The unnoticeable loss is applied on all nodes $\mathcal{V}$ to ensure that the generated trigger meets the unnoticeable constraint for various kinds of nodes.

\vspace{0.2em}
\noindent \textbf{Bi-level Optimization.} To guarantee the effectiveness of the generated triggers, we optimize the adaptive trigger generator to successfully attack the surrogate GCN model $f_s$ with a bi-level optimization. Specifically, the surrogate GCN $f_s$ will be trained on the backdoored dataset, which can be formulated as:
\begin{equation}
    \min_{\theta_s} \mathcal{L}_s(\theta_s, \theta_g) =  \sum_{v_i \in \mathcal{V}_L}l(f_s(\mathcal{G}_C^i), y_i) + \sum_{v_i \in \mathcal{V}_P}l(f_{s}(a(\mathcal{G}_C^i,g_i)), y_t),
    \label{eq:inner}
\end{equation}
where $\theta_s$ represents the parameters of the surrogate GCN $f_s$, $\mathcal{G}_C^i$ indicates the clean computation graph of node $v_i$, and $a(\cdot)$ denotes the attachment operation. $y_i$ is the label of labeled node $v_i \in \mathcal{V}_L$ and $y_t$ is the target class label. The adaptive trigger will be optimized to effectively mislead the surrogate model $f_s$ to predict various nodes from $\mathcal{V}$ to be $y_t$ once injected with adaptive triggers, which can be written as:
\begin{equation}
    \mathcal{L}_g(\theta_s, \theta_g) = \sum_{v_i \in \mathcal{V}}l(f_s(a(\mathcal{G}_C^i,g_i)), y_t).
\end{equation}
Combining the unnoticeable loss Eq.(\ref{eq:l_c}), the following bi-level optimization problem can be formulated:
\begin{equation}
\begin{aligned}
    \min_{\theta_g} & \mathcal{L}_g (\theta_s^*(\theta_g),\theta_g) + \beta \mathcal{L}_c(\theta_g) \\
    s.t. & \quad \theta_s^* = \arg \min_{\theta_s} \mathcal{L}_s(\theta_s,\theta_g),
    \label{eq:bilevel}
\end{aligned}
\end{equation}
where $\beta$ is used to control the contribution of unnoticeable loss.
\subsection{Optimization Algorithm}
We propose an alternating optimization schema to solve the bi-level optimization problem of Eq.(\ref{eq:bilevel}) with a small computation cost. 

\noindent \textbf{Updating Lower Level Surrogate Model.} Computing $\theta_s^*$ for each outter iteration is expensive. We update surrogate model $\theta_s$ with $N$ inner iterations with fixed $\theta_g$ to approximate $\theta_s^*$ as~\cite{zugner2019adversarial} does:
\begin{equation}
    \theta_s^{t+1} = \theta_s^t - \alpha_s \nabla_{\theta_s} \mathcal{L}_s(\theta_s, \theta_g)
    \label{eq:lower_level_sur}
\end{equation}
where $\theta_s^t$ denotes model parameters after $t$ iterations. $\alpha_s$ is the learning rate for training the surrogate model.

\noindent \textbf{Updating Upper Level Surrogate Model.} In the outer iteration, the updated surrogate model parameters $\theta_s^T$ are used to approximate $\theta_s^*$. Moreover, we apply first-order approximation~\cite{finn2017model} in computing gradients of $\theta_g$ to further reduce the computation cost:
\begin{equation}
    \theta_g^{k+1} = \theta_g^k - \alpha_g \nabla_{\theta_g} \big(\mathcal{L}_g(\bar{\theta}_s, \theta_g^k) + \beta \mathcal{L}_c(\theta_g^k)\big),
    \label{eq:upper_level_sur}
\end{equation}
where $\bar{\theta}_s$ indicates gradient propagation stopping. $\alpha_g$ is the learning rate of training adaptive generator. See more details in algorithm~\ref{alg:Framwork}. And the time complexity analysis can be found in Appendix~\ref{app:complexity}.

\section{Experiments}
 In this section, we will evaluate proposed methods on various large-scale datasets to answer the following research questions:
\begin{itemize}[leftmargin=*]
    % \item \textbf{RQ1}: How powerful are our proposed defense methods against existing backdoor attacks? (How easily existing backdoor attacks are defended?)
    \item \textbf{RQ1}: Can our proposed method conduct effective backdoor attacks on GNNs and simultaneously ensure unnoticeability?
    \item \textbf{RQ2}: How do the number of poisoned nodes affect the performance of backdoor attacks?
    \item \textbf{RQ3}: How do the adaptive constraint and the poisoned node selection module affect the attack performance? 
\end{itemize}

\subsection{Experimental Settings}
\subsubsection{Datasets}
To demonstrate the effectiveness of our {\method}, we conduct experiments on four public real-world datasets, i.e., Cora, Pubmed~\cite{sen2008collective}, Flickr~\cite{zeng2020graphsaint}, and OGB-arxiv~\cite{hu2020ogb}, that are widely used for inductive semi-supervised node classification. Cora and Pubmed are small citation networks. Flickr is a large-scale graph that links image captions sharing the same properties. OGB-arixv is a large-scale citation network. The statistics of the datasets are summarized in Tab. \ref{tab:dataset}.

\subsubsection{Compared Methods} \label{sec:baseline}
We compare UGBA with representative and state-of-the-art graph backdoor attack methods, including \textbf{GTA}~\cite{xi2021graph}, \textbf{SBA-Samp}~\cite{zhang2021backdoor} and its variant \textbf{SBA-Gen}. We also compare \textbf{GBAST}~\cite{sheng2021backdoor} on Pubmed, which is shown in the Appendix~\ref{app:add}.

As {\method} conduct attacks by injecting triggers to target nodes, we also compare {\method} with two state-of-the-art graph injection evasion attacks designed for large-scale attacks, i.e. \textbf{TDGIA}~\cite{zou2021tdgia} and \textbf{AGIA}~\cite{chen2022giahao}. More details of these compared methods can be found in Appendix~\ref{app:baseline}.
For a fair comparison, hyperparameters of all the attack methods are tuned based on the performance of the validation set.

\vspace{0.3em}
\noindent \textbf{Competing with Defense Methods.}
We applied the backdoor defense strategies introduced in Sec.~\ref{sec:3.3.2} (i.e., Prune and Prune+LD) to help evaluate the unnoticeability of backdoor attacks. Moreover, two representative robust GNNs, i.e., \textbf{RobustGCN}~\cite{zhu2019robust} and \textbf{GNNGuard}~\cite{zhang2020gnnguard}, are also selected to verify that {\method} can also effectively attack general robust GNNs. 
\subsubsection{Evaluation Protocol} \label{sec:eval}
\begin{figure}[t]
    \small
    \centering
    \begin{subfigure}{0.49\linewidth}
        \includegraphics[width=0.98\linewidth]{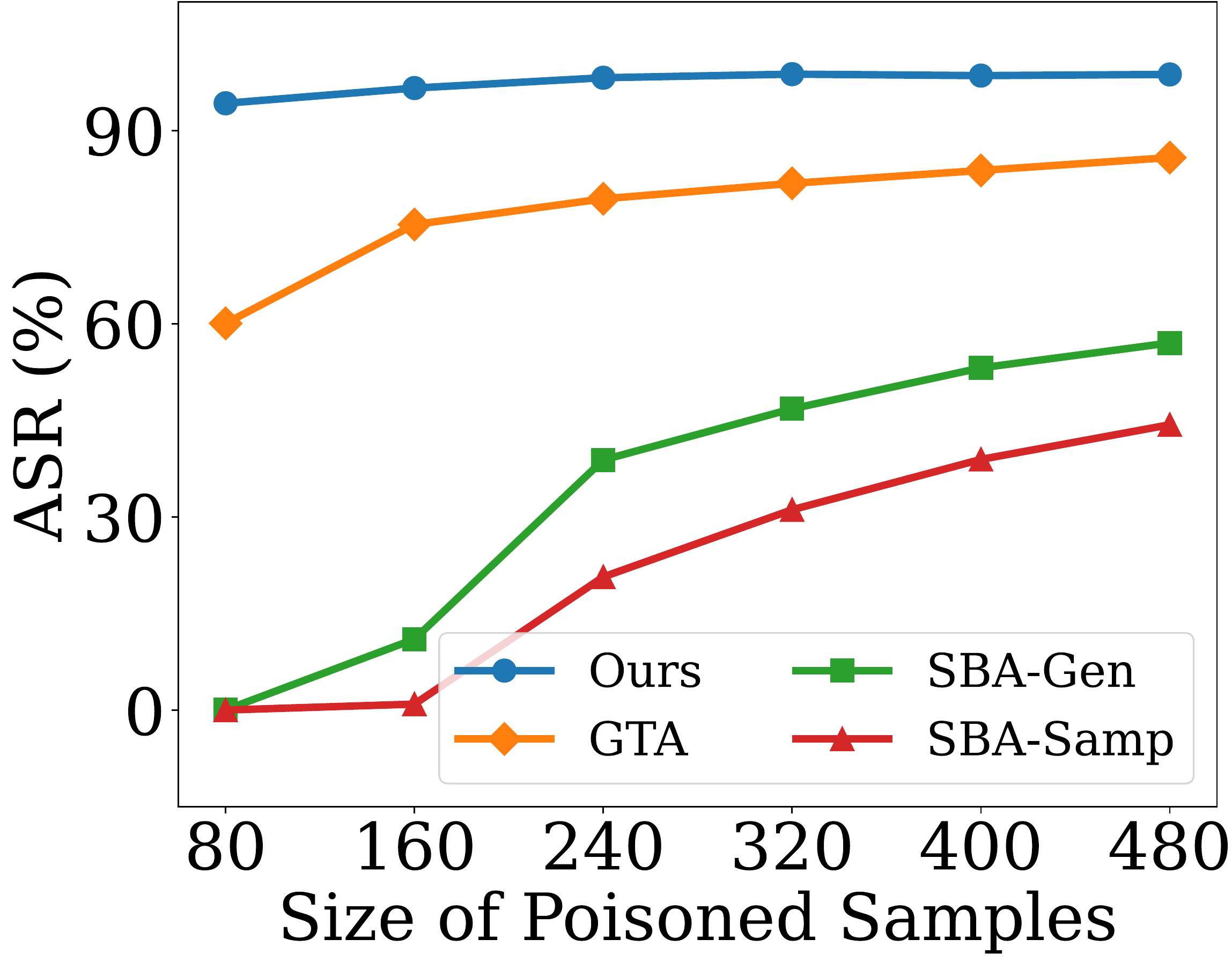}
        \vskip -0.5em
        \caption{No Defense}
    \end{subfigure}
    \begin{subfigure}{0.49\linewidth}
        \includegraphics[width=0.98\linewidth]{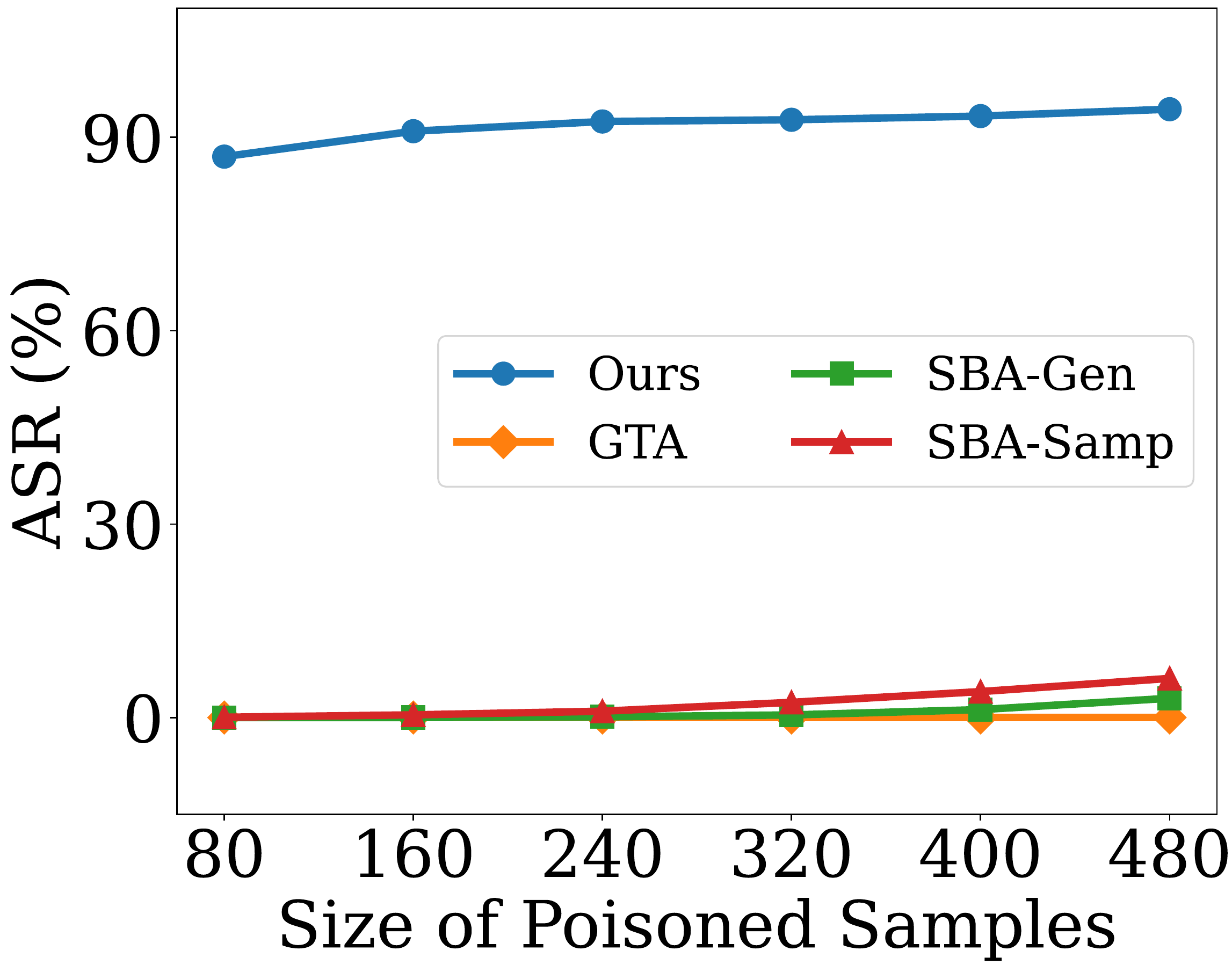}
        \vskip -0.5em
        \caption{Prune+LD}
    \end{subfigure}
    \vskip -1.5em
    \caption{Impacts of sizes of poisoned nodes on OGB-arxiv.}
    \vskip -1em
    \label{fig:poison_size}
\end{figure}

In this paper, we conduct experiments on the inductive node classification task, \textit{where the attackers can not access test nodes when they poison the graph}. Hence, we randomly mask out 20\% nodes from the original dataset. And half of the masked nodes are used as target nodes for attack performance evaluation. Another half is used as clean test nodes to evaluate the prediction accuracy of backdoored models on normal samples. The graph containing the rest 80\% nodes will be used as training graph $\mathcal{G}$, where the labeled node set and validation set both contain 10\% nodes.
The average success rate (ASR) on the target node set and clean accuracy on clean test nodes are used to evaluate the backdoor attacks.
A two-layer GCN is used as the surrogate model for all attack methods. And to demonstrate the transferability of the backdoor attacks, we attack target GNNs with different architectures, i.e., \textbf{GCN}, \textbf{GraphSage}, and \textbf{GAT}. Experiments on each target GNN architecture are conducted 5 times. We report the average ASR and clean accuracy of the total 15 runs (Tab.~\ref{tab:RQ1_table}, Fig.~\ref{fig:abla}, and Fig.~\ref{fig:poison_size}). 
For all experiments, class $0$ is the target class. The attack budget $\Delta_P$ on size of poisoned nodes $\mathcal{V}_P$ is set as 10, 40, 80 and 160 for Cora, Pubmed, Flickr and OGB-arxiv, respectively. The number of nodes in the trigger size is limited to 3 for all experiments. For experiments varying the budget in trigger size, please refer to Appendix~\ref{app:trigger_size}.

\begin{table}[t]
    \centering
    \caption{Dataset Statistics}
    \small
    \vskip -1em 
    \begin{tabularx}{0.8\linewidth}{p{0.18\linewidth}XXXXXX}
    \toprule 
    Datasets & \#Nodes & \#Edges & \#Feature & \#Classes \\%& \#Attack Budget\\
    \midrule
    
    Cora & 2,708 & 5,429 & 1,443 & 7 \\ %& 10\\
    Pubmed & 19,717 & 44,338 & 500 & 3 \\ %& 40\\
    Flickr & 89,250 & 899,756 & 500 & 7 \\ %& 80\\
    OGB-arxiv & 169,343 & 1,166,243 & 128 & 40 \\ %& 160\\
    \bottomrule
    \end{tabularx}
    \vskip -1em
    \label{tab:dataset}
\end{table}

Our {\method} deploys a 2-layer GCN as the surrogate model. A 2-layer MLP is used as the adaptive trigger. More details of the hyperparameter setting can be found in Appendix~\ref{sec:implement_details}.

\begin{table*}[t]
    \centering
    \small
    \caption{ Backdoor attack results (ASR (\%) | Clean Accuracy (\%)). Only clean accuracy is reported for clean graphs.}
    \vskip -1em
    \begin{tabularx}{0.93\linewidth}{p{0.08\linewidth}p{0.08\linewidth}CCCCC}
    \toprule
    Datasets & Defense & Clean Graph & SBA-Samp & SBA-Gen & GTA & Ours  \\
    \midrule
    \multirow{3}{*}{Cora}
    & None & 83.09 & 34.94 | 84.09 & 42.54 | 84.81 & 90.25 | 82.88 & \textbf{96.95 | 83.90}\\
    & Prune & 79.68 & 16.70 | 82.98 & 19.56 | 83.19 & 17.63 | 83.06 & \textbf{98.89 | 82.66}\\
    & Prune+LD & 79.68 & 15.87 | 79.63 & 17.49 | 80.61 & 18.35 | 80.17 & \textbf{95.30 | 79.90}\\
    \midrule
    \multirow{3}{*}{Pubmed}
    & None     & 84.86 & 30.43 | 84.93 & 31.96 | 84.93 & 86.64 | 85.07 & \textbf{92.27} | \textbf{85.06}\\
    & Prune    & 85.09 & 22.10 | 84.90 & 22.13 | 84.86 & 28.10 | 85.05 &\textbf{92.87} | \textbf{85.09} \\
    & Prune+LD & 85.12 & 21.56 | 84.63 & 22.06 | 83.71 & 22.00 | 83.76 &\textbf{93.06} | \textbf{83.75} \\
    \midrule
    \multirow{3}{*}{Flickr}
    & None & 46.40     & \hspace{4pt}0.00 | 47.36 & \hspace{4pt}0.00 | 47.07 & 88.64 | 45.67 & \textbf{97.43 | 46.09} \\
    & Prune & 43.02    & \hspace{4pt}0.00 | 44.01 & \hspace{4pt}0.00 | 43.78 & \hspace{4pt}0.00 | 42.71 & \textbf{90.34 | 42.99} \\
    & Prune+LD & 43.02 & \hspace{4pt}0.00 | 45.03 & \hspace{4pt}0.00 | 45.32 & \hspace{4pt}0.00 | 44.99 & \textbf{96.81 | 42.14} \\
    \midrule
    \multirow{3}{*}{OGB-arxiv}
    & None & 65.50     & \hspace{4pt}0.65 | 65.53 & 11.26 | 65.43 & 75.01 | 65.54 & \textbf{96.59} | \textbf{64.10}\\
    & Prune & 62.16    & \hspace{4pt}0.03 | 63.88 & \hspace{4pt}0.01 | 64.10 & \hspace{4pt}0.01 | 63.97 & \textbf{93.07} | \textbf{62.58}\\
    & Prune+LD & 62.16 & \hspace{4pt}0.16 | 64.15 & \hspace{4pt}0.02 | 63.89 &\hspace{4pt}0.03 | 64.30 & \textbf{90.95} | \textbf{63.19}\\
    % \midrule
    % \multirow{3}{*}{Reddit}
    % & None & 86.17\\
    % & Prune & 87.20\\
    % & Prune+LD & 86.59\\
    \bottomrule
    \end{tabularx}
    \vskip -1em
    \label{tab:RQ1_table}
\end{table*}

\begin{table}[t]
    \centering
    \caption{Comparisons of ASR (\%) with node inject attacks.}
    \small
    \vskip -1em 
    \begin{tabularx}{0.98\linewidth}{Xp{0.18\linewidth}XXX}
    \toprule 
    Datasets & Defense &TDGIA & AGIA & Ours \\
    \midrule
    \multirow{3}{*}{Flickr}
    & GCN-Prune & 77.01 & 77.22 & \textbf{99.91}\\
    & RobustGCN & 78.61 & 78.61 & \textbf{99.23}\\
    & GNNGuard & 55.68 & 56.01 & \textbf{99.91}\\
    \midrule
    \multirow{3}{*}{OGB-arixv}
    & GCN-Prune & 66.17 & 66.33 & \textbf{94.05}\\
    & RobustGCN & 73.87 & 74.00 & \textbf{95.39}\\
    & GNNGuard  & 42.27 & 42.58 & \textbf{96.88}\\
    \bottomrule
    \end{tabularx}
    \vskip -1.5em
    \label{tab:RQ1_evasion}
\end{table}

\subsection{Attack Results}
To answer \textbf{RQ1}, we compare UGBA with baselines on four real-world graphs under various defense settings in terms of attack performance and unnoticeability. 
\subsubsection{Comparisons with baseline backdoor attacks}
We conduct experiments on four real-world graphs under three backdoor defense strategy settings (i.e., No defense, Prune and Prune+LD).
As described by the evaluation protocol in Sec.~\ref{sec:eval}, we report the average results in backdooring three target GNN architectures in Tab.~\ref{tab:RQ1_table}. The details of the backdoor attack results are presented in Tab.~\ref{tab:RQ1_table_Appendix1}-\ref{tab:RQ1_table_Appendix3} in Appendix. From the table, we can make the following observations:
\begin{itemize}[leftmargin=*]
    \item When no backdoor defense strategy is applied, our {\method} outperforms the baseline methods, especially on large-scale datasets. This indicates the effectiveness of poisoned node selection algorithm in fully utilizing the attack budget. 
    \item All the baselines give poor performance when the trigger detection based defense methods, i.e., Prune and Prune+LD, are adopted. By contrast, our {\method} can achieve over 90\% ASR with the defense strategies and maintain high clean accuracy. This demonstrates that our {\method} can generate effective and unnoticeable triggers for backdoor attacks. 
    \item As the ASRs are average results of backdooring three different GNN architectures, the high ASR scores of UGBA prove its transferability in backdooring various types of GNN models. 
\end{itemize}

\subsubsection{Comparisons with baseline node injection attacks}
We also compare UGBA with two state-of-the-art node injection evasion attacks. Experiments are conducted on Flickr and OGB-arxiv. Three defense models (GCN-Prune, RobustGCN and GNNGuard) are selected to defend against the compared attacks. The ASR of 5 runs is reported in Tab~\ref{tab:RQ1_evasion}. From this table, we observe:
\begin{itemize}[leftmargin=*]
    \item {\method} can effectively attack the robust GNNs, which shows that {\method}  can also bypass the general defense methods with the unnoticeable constraint.
    \item Compared with node injection attacks, {\method} only requires a very small additional cost in injecting triggers and labels (e.g. 160 poisoned nodes out of 169K nodes in OGB-arxiv). But  {\method} can outperform node injection attacks by 30\%.  This implies the superiority of {\method} in attacking large amounts of target nodes.
\end{itemize}

\subsection{Impacts of the Sizes of Poisoned Nodes}
To answer \textbf{RQ2}, we conduct experiments to explore the attack performance of {\method} given different budgets in the size of poisoned nodes. Specifically, we vary the sizes of poisoned samples as $\{80,160,240,320,400,480\}$. The other settings are the same as the evaluation protocol in Sec.~\ref{sec:eval}.
Hyperparameters are selected with the same process as described in Appendix. \ref{sec:implement_details}. 
Fig. \ref{fig:abla} shows the results on OGB-arxiv. We have similar observations on other datasets. From Fig. \ref{fig:abla}, we can observe that:
\begin{itemize}[leftmargin=*]
    \item The attack success rate of all compared methods in all settings increases as the increase of the number of poisoned samples, which satisfies our expectation. Our method consistently outperforms the baselines as the number of poisoned samples increases, which shows the effectiveness of the proposed framework. In particular, the gaps between our method and baselines become larger when the budget is smaller, which demonstrates the effectiveness of the poisoned node selection in effectively utilizing the attack budget. 
    % \item When no defense is applied on the backdoor attack, our method can achieve over $90\%$ ASR with only 80 poisoned samples (out of 169,343 nodes) used in the attack on OGB-arxiv. This is because our poisoned node selection algorithm selects the most representative nodes on the graphs as the poisoned nodes, which enforces other nodes are also conducted successful backdoor attacks. Therefore, we can conclude that our approach can use the least amount of budget to achieve the best attack performance.
    \item When Prune+LD defense is applied on the backdoor attacks, our methods still achieve promising performances, while all the baselines obtain nearly $0\%$ ASR in all settings, which is as expected. That's because our method can generate trigger nodes similar to the attached nodes due to the unnoticeable constraint, which is helpful for bypassing the defense method.
\end{itemize}

\subsection{Ablation Studies}
\begin{figure}[t]
    \small
    \centering
    \begin{subfigure}{0.49\linewidth}
        \includegraphics[width=0.98\linewidth]{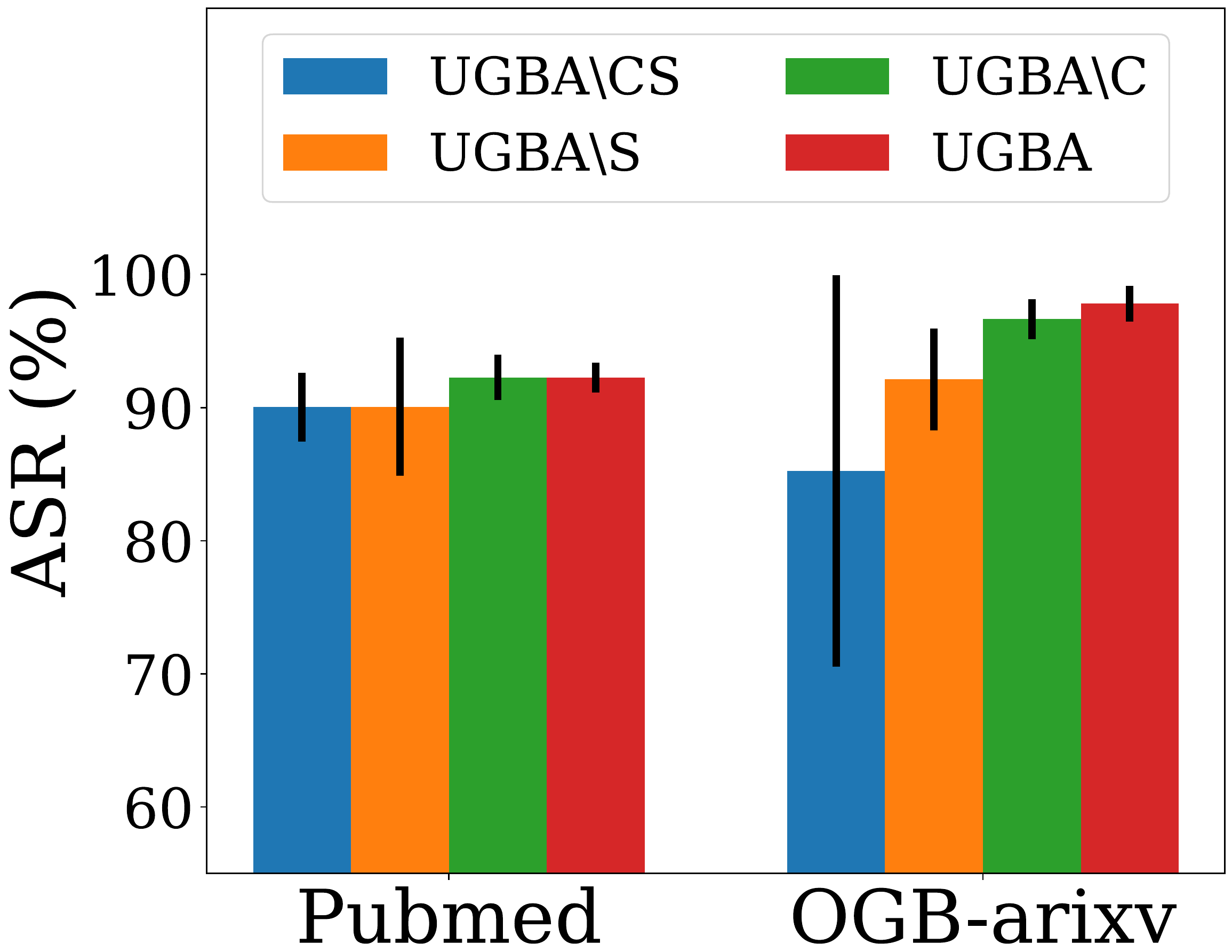}
        \vskip -0.5em
        \caption{No defense}
    \end{subfigure}
    \begin{subfigure}{0.49\linewidth}
        \includegraphics[width=0.98\linewidth]{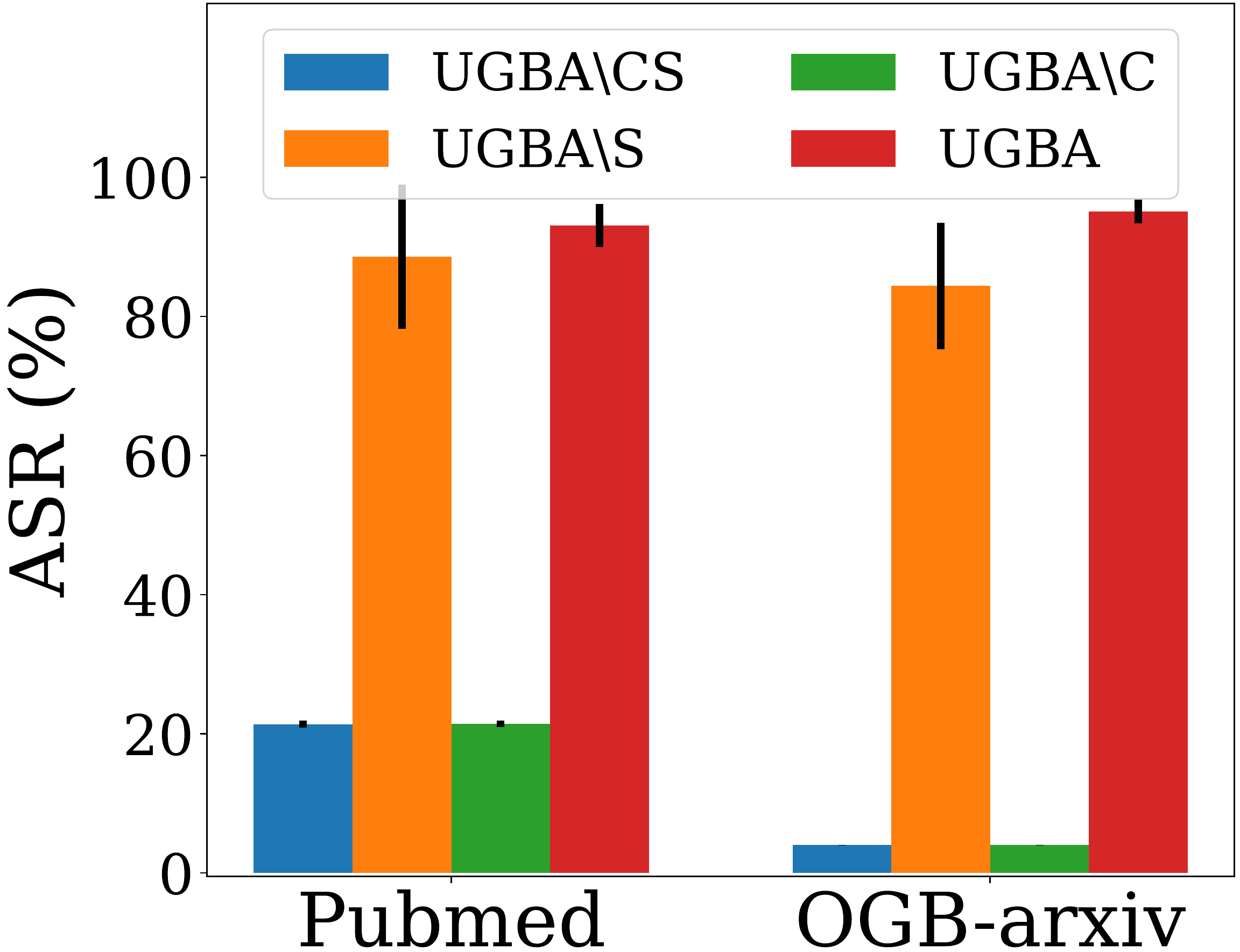}
        \vskip -0.5em
        \caption{Prune+LD}
    \end{subfigure}
    \vskip -1.2em
    \caption{Ablation studies on Pubmed and OGB-arxiv.}
    \vskip -1em
    \label{fig:abla}
\end{figure}

To answer \textbf{RQ3}, we conduct ablation studies to explore the effects of the unnoticeable constraint and the poisoned node selection module. To demonstrate the effectiveness of the unnoticeable constraint module, we set the $\beta$ as 0 when we train the trigger generator and obtain a variant named as {\method}$\backslash$C. To show the benefits brought by our poisoned node selection module, we train a variant {\method}$\backslash$S which randomly selects poisoned nodes to attach triggers and assign target nodes. We also implement a variant of our model by removing both unnoticeable constraint and poisoned node selection, which is named as {\method}$\backslash$CS.
The average results and standard deviations on Pubmed and OGB-arxiv are shown in Fig.~\ref{fig:abla}. All the settings of evaluation follow the description in Sec.~\ref{sec:eval}. And the hyperparameters of the variants are also tuned based on the validation set for fair comparison. From Fig.~\ref{fig:abla}, we observe that: 
\begin{itemize}[leftmargin=*]
    \item Compared with {\method}$\backslash$S, {\method} achieves better attack results on various defense settings. The variance of ASR of {\method} is significantly lower than that of {\method}$\backslash$S. This is because our poisoned node selection algorithm selects consistently diverse and representative nodes that are useful for backdoor attacks.
    \item When the backdoor defense strategy Prune+LD, {\method} can outperform {\method}$\backslash$C and {\method}$\backslash$CS by a large margin. This implies that the proposed unnoticeable loss manages to guide the trigger generator to give unnoticeable triggers for various test nodes, which can effectively bypass the pruning defenses. 
\end{itemize}

\begin{figure}[h]
    \small
    \centering
    \vskip -0.5em
    \begin{subfigure}{0.49\linewidth}
        \includegraphics[width=0.9\linewidth]{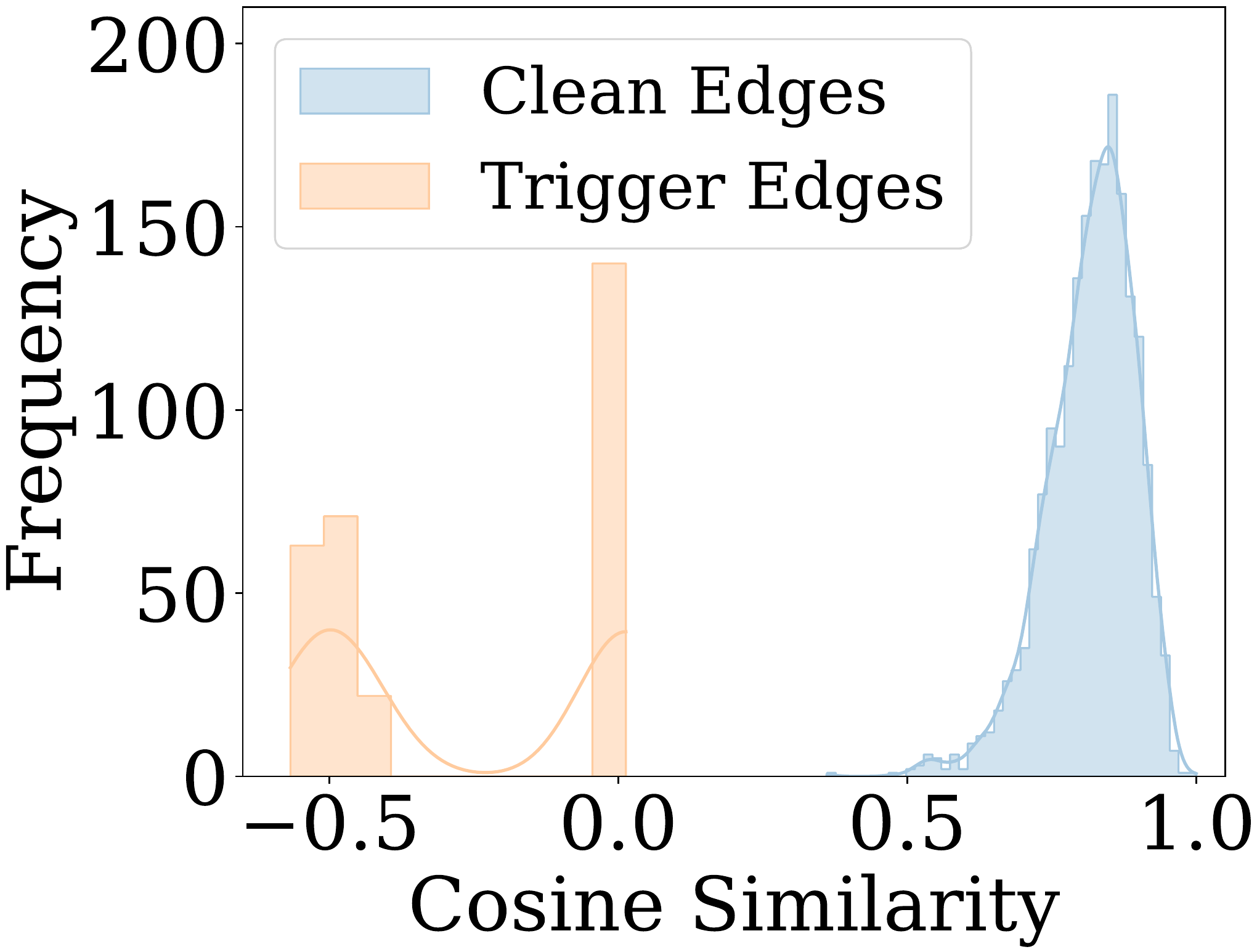}
        \vskip -0.5em
        \caption{GTA}
    \end{subfigure}
    \begin{subfigure}{0.49\linewidth}
        \includegraphics[width=0.9\linewidth]{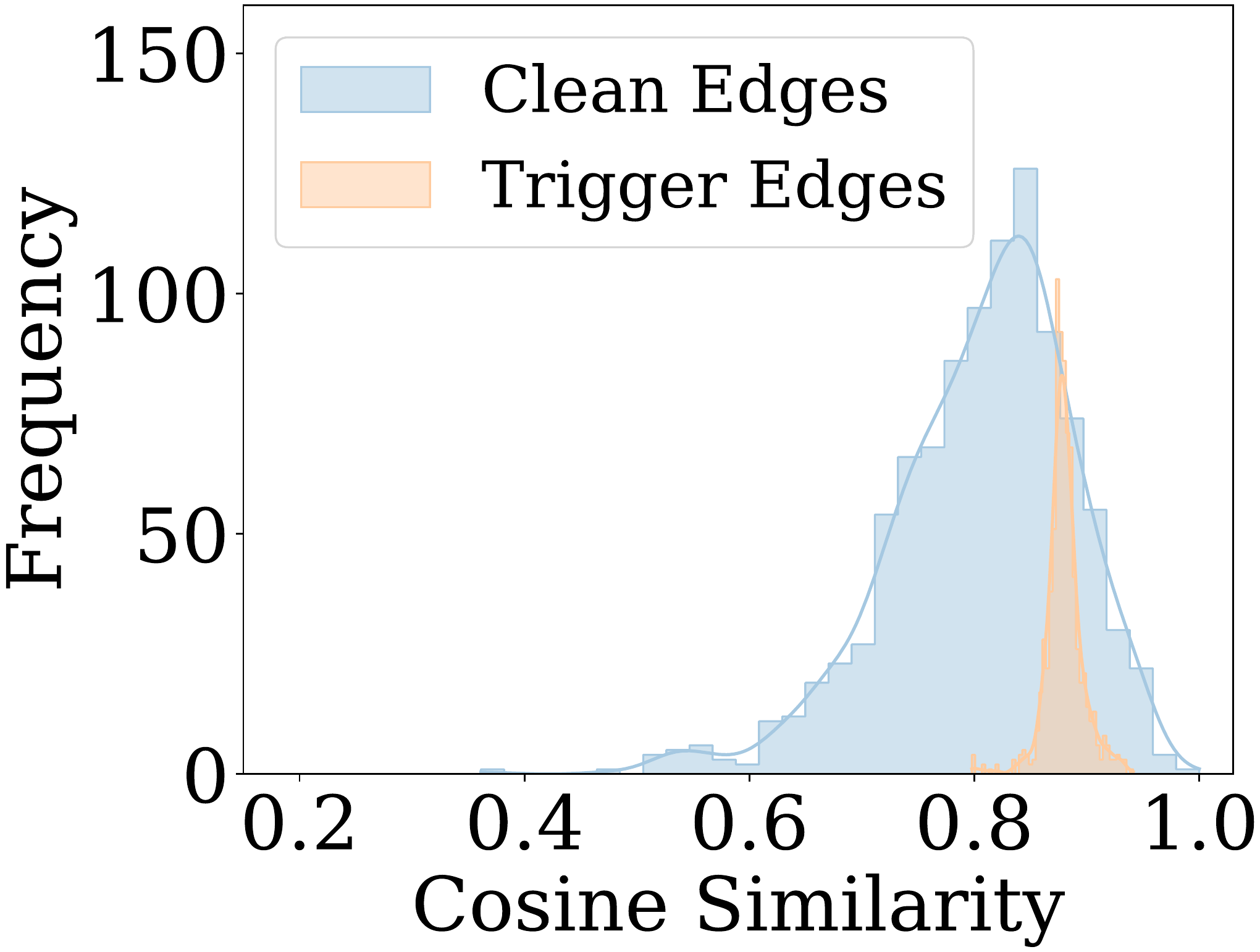}
        \vskip -0.5em
        \caption{UGBA}
    \end{subfigure}
    \vskip -1em
    \caption{Edge similarity distributions on OGB-arxiv.}
    \vskip -1.5em
    \label{fig:sim}
\end{figure}
\subsection{Similarity Analysis}
In this section, we conduct a case study to further explore the similarity of the trigger nodes. We conduct backdoor attacks by using both GTA and our method on OGB-arxiv and then calculate the edge similarities of trigger edges (i.e., the edges associated with trigger nodes) and clean edges (i.e., the edges not connected to trigger nodes). The histogram of the edge similarity scores are plotted in Fig. \ref{fig:sim}. From the figure, we observe that the trigger edges generated by GTA have low similarities, which implies high risk of trigger elimination with our proposed backdoor defense strategies. In contrast, the edges created by our method present cosine similarity scores that well disguise them as clean edges, which verifies the unnoticeability of our methods.

\subsection{Parameter Sensitivity Analysis}
\begin{figure}[t]
    \small
    \centering
    \begin{subfigure}{0.49\linewidth}
        \includegraphics[width=0.9\linewidth]{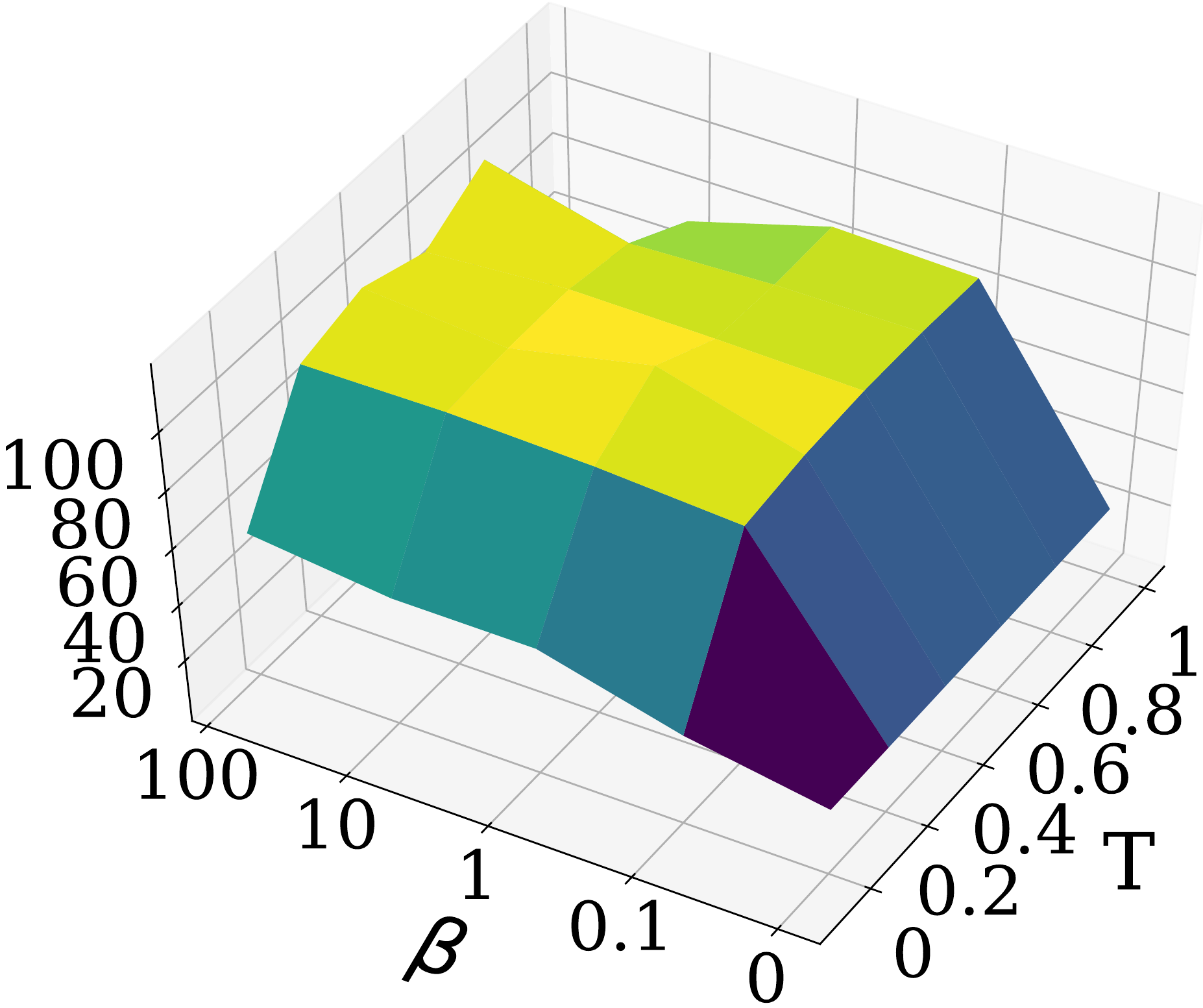}
        \vskip -0.5em
        \caption{Pubmed}
    \end{subfigure}
    \begin{subfigure}{0.49\linewidth}
        \includegraphics[width=0.9\linewidth]{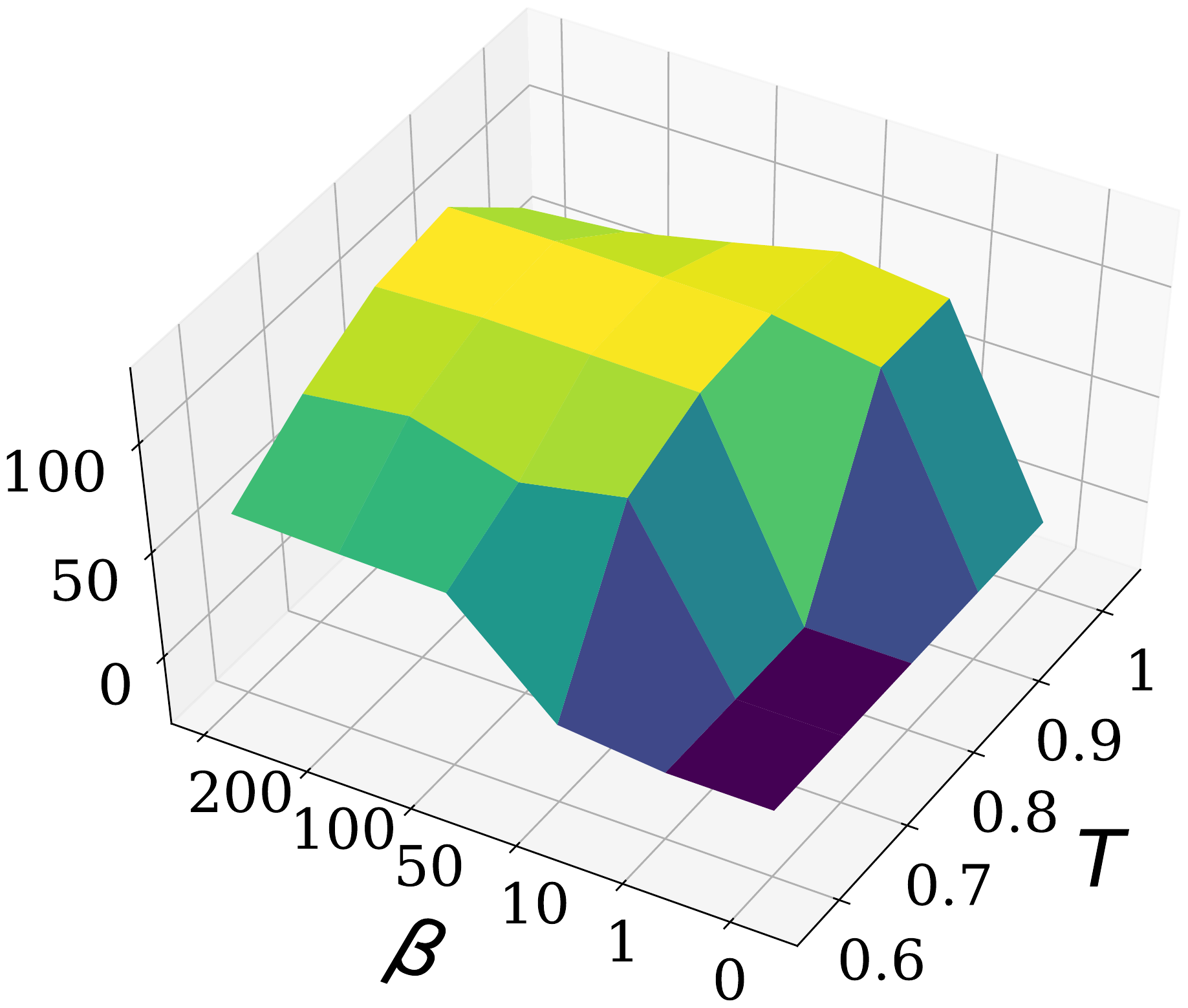}
        \vskip -0.5em
        \caption{OGB-arxiv}
    \end{subfigure}
    \vskip -1.5em
    \caption{Hyperparameter Sensitivity Analysis}
    \vskip -1.5em
    \label{fig:hyper}
\end{figure}

In this subsection, we further investigate how the hyperparameter $\beta$ and $T$ affect the performance of UGBA, where $\beta$ and $T$ control the weight of unnoticeable loss in training the trigger generator and the threshold of similarity scores used in unnoticeable loss. To explore the effects of $\beta$ and $T$, we vary the values of $\beta$ as $\{0,50,100,150,200\}$. And $T$ is changed from $\{0,0.2,0.4,0.6,0.8,1\}$ and $\{0.6,0.7,0.8,0.9,1\}$ for Pubmed and OGB-arxiv, respectively. Since $\beta$ and $T$ only affect the unnoticeablity of triggers, we report the attack success rate (ASR) of attacking against the Prune+LD defense strategy in Fig.~\ref{fig:hyper}. The test model is fixed as GCN. 
We observe that (i): In Pubmed, the similarity threshold $T$ needs to be larger than 0.2; while $T$ is required to be higher than 0.8 in OGB-arxiv. This is because edges in OGB-arxiv show higher similarity scores compared with Pubmed. Hence, to avoid being detected, a higher similarity threshold $T$ is necessary. In practice, the $T$ can be set according to the average edge similarity scores of the dataset.
(ii) When $T$ is set to a proper value, high ASR can generally be achieved when $\beta \leq 1$, which eases the hyperparameter tuning.

% As the values of $T$ and $\beta$ increase, the ASR of our method also gradually increases. However, if the values of $\beta$ and $T$ are too large, the performance would decrease; (ii): Using low $\beta$ and $T$ would lead to poor performances. We analyze the reason is that a low $T$ will degrade the capability of the trigger generator to learn to generate trigger nodes similar to clean nodes, and a low $\beta$ will limit the upper bound of the similarities between trigger nodes and clean nodes. Thus, the defenders can easily diminish the impacts of triggers by pruning the dissimilar edges. (iii): A $T$ between 0.2 to 0.8 and a $\beta$ between 0.1 to 100 will give good performances to UGBA in Pubmed. As for OGB-arxiv, the choices of $T$ and $\beta$ are about 0.8 and between 10 to 200, respectively. Therefore, our method will generally have great performance in both datasets when $T$ is 0.8 and $\beta$ is between 10 to 100, which simplifies the parameter selection. 

\section{Conclusion and Future Work}
In this paper, we empirically verify that existing backdoor attacks require large attack budgets and can be easily defended with edge pruning strategies. To address these problems, we study a novel problem of conducting unnoticeable graph backdoor attacks with limited attack budgets. Specifically, a novel poisoned node selection algorithm is adopted to select representative and diverse nodes as poisoned nodes to fully utilize the attack budget. And an adaptive generator is optimized with an unnoticeable constraint loss to ensure the unnoticeability of generated triggers. The effectiveness of generated triggers is further guaranteed by bi-level optimization with the surrogate GCN model. Extensive experiments on large-scale datasets demonstrate that our proposed method can effectively backdoor various target GNN models and even be adopted with defense strategies. There are two directions that need further investigation. First, in this paper, we only focus on node classification. We will extend the proposed attack to other tasks such as recommendation and graph classification. Second, it is also interesting to investigate how to defend against the unnoticeable graph backdoor attack.
\newpage
\begin{acks}
This material is based upon work supported by, or in part by, the National Science Foundation (NSF) under grant number IIS-1707548 and IIS-1909702, the Army Research Office (ONR) under grant number W911NF21-1-0198, and Department of Homeland Security (DNS) CINA under grant number E205949D. The findings in this paper do not necessarily reflect the view of the funding agency. 
\end{acks}

%% The file named.bst is a bibliography style file for BibTeX 0.99c
\bibliographystyle{ACM-Reference-Format}
\bibliography{ref}
\appendix
\newpage

% \section{Details of Backdoor Attack Results}
\section{Training Algorithm}
The algorithm of UGBA is proposed in Algorithm $\ref{alg:Framwork}$. Specially, we first select the poisoned nodes $\mathcal{V}_P$ with the top-$n$ highest scores $m(\cdot)$ based on Eq. (\ref{eq:node_sel_score}), and assign the target class $y_t$ as labels to $\mathcal{V}_P$ (lines 3-4). From line 5 to line 10, we train the trigger generator $f_g$ on the surrogate GCN $f_s$ by solving a bi-level optimization problem based on Eq. (\ref{eq:bilevel}). In detail, we update lower level surrogate model (lines 6-8) and upper level surrogate model (line 9), respectively, by doing gradient descent on $\theta_s$ and $\theta_g$ based on Eq.~(\ref{eq:lower_level_sur}) and Eq.~(\ref{eq:upper_level_sur}). After that, from line 11 to line 14, we use the well-trained $f_g$ to generate a trigger $g_i$ for each poisoned node $v_i \in \mathcal{V}_P$ and 
attach $g_i$ with $v_i$ to obtain the poisoned graph $\mathcal{G}_B$.
\begin{algorithm}[t!] 
\caption{Algorithm of {\method}.} 
\label{alg:Framwork} 
\begin{algorithmic}[1]
\REQUIRE $\mathcal{G}=(\mathcal{V}, \mathcal{E}, \mathbf{X})$, $\mathcal{Y}_L$, $\beta$, $T$.
\ENSURE Backdoored dataset $\mathcal{G}_B$, adaptive trigger generator $f_{g}$
% \STATE Calculate the score $m(\cdot)$ based on Eq~(\ref{eq:node_sel_score}) for $\mathcal{V}_U=\mathcal{V}-\mathcal{V}_L-\mathcal{V}_T$
% \ref{eq:node_sel_score} for the unlabeled nodes $\mathcal{V}_{P}$
\STATE Initialize $\mathcal{G}_B = \mathcal{G}$;
\STATE Randomly initialize $\theta_s$ and $\theta_g$ for $f_s$ and $f_g$;
\STATE Select poisoned nodes $\mathcal{V}_P$ based on Eq. (\ref{eq:node_sel_score});
\STATE Assign class $t$ as labels of $\mathcal{V}_{P}$;
% \COMMENT Solve the bi-level optimization
\WHILE{not converged yet}
    % \COMMENT{Update lower level surrogate model}
    \FOR{t=$1,2,\dots{},N$}
    \STATE Update $\theta_s$ by descent on  $\nabla_{\theta_s} \mathcal{L}_s$ based on Eq. (\ref{eq:lower_level_sur});
    \ENDFOR
    \STATE Update $\theta_g$ by descent on $\nabla_{\theta_g} \big(\mathcal{L}_g + \beta \mathcal{L}_c\big)$ based on Eq. (\ref{eq:upper_level_sur});
\ENDWHILE
\FOR{$v_i \in \mathcal{V}_P$}
    \STATE Generate the trigger $g_i$ for $v_i$ by using $f_g$;
    % \STATE Train an adaptive trigger $f_{g}$ by solving Eq. (?) on a surrogate GCN and generate the triggers $g_t$.
    \STATE Update $\mathcal{G}_B$ based on $a(\mathcal{G}_B^i,g_i)$;
    
\ENDFOR
\RETURN $\mathcal{G}_B$ and $f_{g}$;
\label{algorithm}
\end{algorithmic}
\end{algorithm}

\section{Implementation Details}
\label{sec:implement_details}
A 2-layer GCN is deployed as the surrogate model. A 2-layer MLP is used as the adaptive trigger. All the hidden dimension is set as 32.  The inner iterations step $N$ is set as 5 for all the experiments. For the hyperparameter $\beta$ and $T$, they are selected based on the grid search on the validation set. 
Specifically, $T$ is fixed as 0.5, 0.5, 0.5, 0.8 for Cora, Pubmed, Flickr and OGB-arxiv, respectively.  For Prune and Prune+LD defenses, the threshold of pruning is set to filter out around 10\% dissimilar edges. In particular, the set thresholds are around 0.1, 0.2, 0.4, 0.8 for Cora, Pubmed, Flickr and OGB-arxiv.

% \section{Details of Datasets} \label{app:dataset}
% Here, we present the details of the datasets we used.
% \begin{itemize}[leftmargin=*]
%     \item Cora and Pubmed~\cite{sen2008collective}: 
%     Widely-used benchmarks for node classification. Each node represents a machine-learning papers and is divided into one of seven classes. Each edge indicates that one paper cites another one. Each paper is described by a 0/1-valued word vector indicating the absence/presence of the corresponding word from the dictionary. 
%     \item Flickr~\cite{zeng2020graphsaint}: A large-scale graph that contains descriptions and common properties of images to categorize the type of images. Each node represents a image and edges are formed between nodes sharing common properties. The node features contains information of low-level feature from ~\cite{chua2009nuswide}.\
%     \item OGB-arxiv~\cite{hu2020ogb}: A large-scale open benchmark graph representing the citation network between all computer science arxiv papers. Each node is an arxiv paper with a 128-dimensional feature vector obtained by averaging the embeddings of words in its title and abstract.
% \end{itemize}

\section{Additional Experiments}
\label{app:add}
We compare our {\method} with GBAST~\cite{sheng2021backdoor} on Pubmed. And we report the ASR (\%) of attacking GCN under different defense settings in Tab.~\ref{tab:add}. We similar observations on other datasets and target GNN models. Compared with GBAST which selects the poisoned samples by degree and closeness centrality, our {\method} achieves much higher ASR when not defense is applied. This implies the effectiveness of our clustering-based poisoned sample selection. In addition, GBAST cannot bypass the defense strategy. Our {\method} can still show high attack performance under the Prune+LD defense.
\begin{table}[h]
    \centering
    \small
    \caption{Comparison with GBAST}
    \vskip -1.5em
    \begin{tabularx}{0.7\linewidth}{lCC}
    \toprule
    Defense     & GBAST & {\method} \\
    \midrule
    None        & 55.1 $\pm 11.4$ & 96.3 $\pm 1.3$ \\
    Prune+LD    & 21.3 $\pm 0.2$ & 92.3 $\pm 2.1$ \\
    \bottomrule
    \end{tabularx}
    \vskip -1em
    \label{tab:add}
\end{table}

\section{Details of Compared Methods} \label{app:baseline}
The details of compared methods are described following:
\begin{itemize}[leftmargin=*]
    \item \textbf{SBA-Samp}~\cite{zhang2021backdoor}: It injects one fixed subgraph as a trigger to the training graph for a poisoned node. To generate the subgraph the connections are generated using Erdos-Renyi (ER) model and the node features are randomly sampled from the training graph.
    \item \textbf{SBA-Gen}: This is a variant of SBA-Samp, which uses generated features for trigger nodes. Features are from a Gaussian distribution whose mean and variance is computed from real nodes.
    \item \textbf{GTA}~\cite{xi2021graph}: This is the state-of-the-art backdoor attack on GNNs. Poisoned nodes is randomly selected in GTA. A trigger generator is adopted to create subgraphs as sample-specific triggers. The trigger generator is purely optimized by the backdoor attack loss without any unnoticeable constraint. 
    \item \textbf{TDGIA}~\cite{zou2021tdgia}: It employs a topological defective edge selection strategy to choose the nodes to be connecting with the injected ones, and generates the features for injected nodes by performing the smooth adversarial feature optimization.
    \item \textbf{AGIA}~\cite{chen2022giahao}: It leverages gradient information to perform a bi-level optimization for the features and structures of the injected nodes.
\end{itemize}

\section{Impacts of trigger size} \label{app:trigger_size}
In this section, we conduct experiments to explore the attack performance of UGBA by injecting different numbers of nodes as a trigger for a poisoned node. Specially, the trigger size is varied as $\{1,2,3,4,5\}$. The other settings are the same as the evaluation protocol in Sec.~\ref{sec:eval}. The results on OGB-arxiv are shown in 
Table~\ref{tab:Size_Appendix}. We have similar observations on other datasets. From the table, we  can find that: (i) as the increase of trigger sizes, the attack success rate of UGBA in all settings increases, as larger trigger can be stronger in backdoor attack; (ii) {\method} can achieve stable and high attack performance when the trigger size is as small as 2, which shows the effectiveness of our {\method} in generating triggers.

\begin{table}[h!]
    \centering
    \caption{Attack results of UGBA (ASR (\%)) with various trigger sizes under three defense strategies on OGB-arxiv }
    \small
    \vskip -1em 
    \begin{tabularx}{0.95\linewidth}{cCCCCCC}
    \toprule 
    \multirow{1}{*}{Trigger Size} & 1 & 2 & 3 & 4 & 5 \\
    \midrule
    None & 83.0 & 98.9 & 98.8 & 98.9 & 97.8\\ 
    Prune & 77.7 & 94.5 & 94.4 & 94.6 & 93.8\\
    Prune+LD & 65.2 & 94.0 & 95.1 & 94.2 & 89.0\\
    \bottomrule
    \end{tabularx}
    \vskip -1em
    \label{tab:Size_Appendix}
\end{table}

\begin{table*}[h!]
    \centering
    \scriptsize
    \caption{ Results of backdooring GCN (ASR (\%) | Clean Accuracy (\%)). Only clean accuracy is reported for clean graph.}
    \vskip -1em
    \begin{tabularx}{0.85\linewidth}{p{0.08\linewidth}p{0.04\linewidth}CCCCC}
    \toprule
    Datasets & Defense & Clean Graph & SBA-Samp & SBA-Gen & GTA & Ours  \\
    \midrule
    \multirow{3}{*}{Cora}
    & None & 82.9 & 33.8$\pm$3.4 | 83.9$\pm$1.1 & 40.1$\pm$7.0 | 83.5$\pm$1.1 & 98.9$\pm$0.8 | 82.7$\pm$1.2 & \textbf{98.8$\pm$0.1} | \textbf{83.5$\pm$0.8}\\
    & Prune & 79.6 & 17.1$\pm$2.3 | 83.1$\pm$1.1 & 19.9$\pm$2.6 | 83.3$\pm$0.9 & 17.7$\pm$3.2 | 83.6$\pm$0.6 & \textbf{99.6$\pm$0.0} | \textbf{82.5$\pm$0.9}\\
    & Prune+LD & 79.6 & 16.6$\pm$2.7 | 81.3$\pm$1.2 & 18.9$\pm$3.5 | 81.3$\pm$1.0 & 20.1$\pm$5.6 | 80.8$\pm$0.4 & \textbf{99.6$\pm$0.1} | \textbf{81.5$\pm$0.6}\\
    \midrule
    \multirow{3}{*}{Pubmed}
    & None     & 85.1 & 26.4$\pm$2.9 | 84.9$\pm$0.2 & 28.8$\pm$3.5 | 85.0$\pm$0.2 & 92.8$\pm$2.9 | 85.2$\pm$0.2 & \textbf{96.3$\pm$1.2} | \textbf{84.9$\pm$0.1}\\
    & Prune    & 85.1 & 22.4$\pm$1.0 | 85.3$\pm$0.1 & 22.6$\pm$0.9 | 85.2$\pm$0.1 & 28.8$\pm$1.2 | 85.1$\pm$1.0 &\textbf{93.0$\pm$1.1} | \textbf{85.4$\pm$0.2} \\
    & Prune+LD & 85.1 & 21.7$\pm$0.9 | 85.0$\pm$0.2 & 22.2$\pm$1.2 | 84.2$\pm$0.3 & 22.3$\pm$0.5 | 84.2$\pm$0.1 &\textbf{92.3$\pm$2.0} | \textbf{85.0$\pm$0.1} \\
    \midrule
    \multirow{3}{*}{Flickr}
    & None & 45.5     & \hspace{9pt}0$\pm$0.0 | 46.7$\pm$0.3 & \hspace{9pt}0$\pm$0.0 | 46.4$\pm$0.1 & 99.9$\pm$0.1 | 45.0$\pm$0.3 & \textbf{96.9$\pm$2.3} | \textbf{44.8$\pm$0.4} \\
    & Prune & 42.3    & \hspace{9pt}0$\pm$0.0 | 44.1$\pm$0.2 & \hspace{9pt}0$\pm$0.0 | 43.4$\pm$0.4 & \hspace{9pt}0$\pm$0.0|  41.7$\pm$0.2 & \textbf{99.9$\pm$0.0} | \textbf{41.7$\pm$0.4} \\
    & Prune+LD & 42.3 & \hspace{9pt}0$\pm$0.0 | 45.6$\pm$0.2 & \hspace{9pt}0$\pm$0.0 | 45.6$\pm$0.2 & \hspace{9pt}0$\pm$0.0 | 44.5$\pm$0.4 & \textbf{96.6$\pm$1.6} | \textbf{44.8$\pm$0.1} \\
    \midrule
    \multirow{3}{*}{OGB-arxiv}
    & None & 65.6     & \hspace{3pt}0.3$\pm$0.1 | 65.8$\pm$0.1 & \hspace{3pt}1.8$\pm$2.4 | 65.8$\pm$0.2 & 75.2$\pm$1.4 | 65.8$\pm$0.1 & \textbf{98.8$\pm$0.1} | \textbf{63.9$\pm$0.5}\\
    & Prune & 62.1    & \hspace{3pt}0.1$\pm$0.1 | 64.5$\pm$0.5 & \hspace{3pt}0.1$\pm$0.1 | 64.0$\pm$0.1 & \hspace{4pt}0.1$\pm$0.1 | 64.0$\pm$0.1 & \textbf{94.0$\pm$0.3} | \textbf{62.2$\pm$0.7}\\
    & Prune+LD & 62.1 & \hspace{3pt}0.1$\pm$0.1 | 64.7$\pm$0.1 & \hspace{3pt}0.1$\pm$0.1 | 64.6$\pm$0.1 &\hspace{4pt}0.1$\pm$0.1 | 64.7$\pm$0.2 & \textbf{93.5$\pm$0.2} | \textbf{63.0$\pm$0.4}\\
    \bottomrule
    \end{tabularx}
    \vskip -1em
    \label{tab:RQ1_table_Appendix1}
\end{table*}

\begin{table*}[h!]
    \centering
    \scriptsize
    \caption{ Results of backdooring GraphSage (ASR (\%) | Clean Accuracy (\%)). Only clean accuracy is reported for clean graph.}
    \vskip -1em
    \begin{tabularx}{0.85\linewidth}{p{0.08\linewidth}p{0.04\linewidth}CCCCC}
    \toprule
    Datasets & Defense & Clean Graph & SBA-Samp & SBA-Gen & GTA & Ours  \\
    \midrule
    \multirow{3}{*}{Cora}
    & None & 81.8 & 34.2$\pm$4.0 | 83.0$\pm$1.5 & 40.4$\pm$5.6 | 82.7$\pm$1.2 & 99.5$\pm$0.4 | 81.3$\pm$1.0 & \textbf{92.7$\pm$2.1} | \textbf{82.8$\pm$1.4}\\
    & Prune & 77.9 & 17.3$\pm$1.8 | 82.5$\pm$1.1 & 19.9$\pm$3.6 | 82.4$\pm$1.1 & 18.6$\pm$4.1 | 81.9$\pm$0.8 & \textbf{99.6$\pm$0.1} | \textbf{83.9$\pm$1.5}\\
    & Prune+LD & 77.9 & 16.4$\pm$1.8 | 78.8$\pm$0.6 & 16.9$\pm$3.2 | 78.9$\pm$0.8 & 18.0$\pm$4.5 | 77.9$\pm$2.0 & \textbf{94.4$\pm$2.1} | \textbf{77.8$\pm$0.5}\\
    \midrule
    \multirow{3}{*}{Pubmed}
    & None     & 85.7 & 38.0$\pm$3.8 | 85.8$\pm$0.3 & 40.0$\pm$4.2 | 85.9$\pm$0.2 & 92.7$\pm$3.5 | 86.0$\pm$0.3 & \textbf{96.0$\pm$0.9} | \textbf{86.0$\pm$0.1}\\
    & Prune    & 86.2 & 22.8$\pm$1.0 | 85.8$\pm$0.2 & 23.0$\pm$0.9 | 85.8$\pm$0.1 & 27.4$\pm$1.2 | 86.5$\pm$0.3 &\textbf{91.0$\pm$0.6} | \textbf{86.4$\pm$0.1} \\
    & Prune+LD & 86.2 & 21.5$\pm$1.0 | 85.4$\pm$0.2 & 22.0$\pm$1.2 | 83.8$\pm$0.1 & 21.9$\pm$0.3 | 83.8$\pm$0.2 &\textbf{91.6$\pm$1.7} | \textbf{86.0$\pm$0.1} \\
    \midrule
    \multirow{3}{*}{Flickr}
    & None & 47.0     & \hspace{9pt}0$\pm$0.0 | 48.5$\pm$0.1 & \hspace{9pt}0$\pm$0.0 | 48.4$\pm$0.1 & 99.7$\pm$0.2 | 48.0$\pm$0.3 & \textbf{98.9$\pm$0.3} | \textbf{47.7$\pm$0.1} \\
    & Prune & 45.2    & \hspace{9pt}0$\pm$0.0 | 46.7$\pm$0.1 & \hspace{9pt}0$\pm$0.0 | 46.8$\pm$0.1 & \hspace{9pt}0$\pm$0.0 | 45.9$\pm$0.2 & \textbf{98.9$\pm$0.9} | \textbf{41.2$\pm$1.3} \\
    & Prune+LD & 45.2 & \hspace{9pt}0$\pm$0.0 | 44.4$\pm$0.4 & \hspace{9pt}0$\pm$0.0 | 44.4$\pm$0.4 & \hspace{9pt}0$\pm$0.0 | 44.5$\pm$0.4 & \textbf{97.1$\pm$2.4} | \textbf{44.7$\pm$0.3} \\
    \midrule
    \multirow{3}{*}{OGB-arxiv}
    & None & 65.6     & \hspace{3pt}0.5$\pm$0.6 | 65.4$\pm$0.6 & \hspace{3pt}6.2$\pm$3.5 | 65.3$\pm$0.6 & 55.5$\pm$2.3 | 65.9$\pm$0.3 & \textbf{91.0$\pm$0.8} | \textbf{63.6$\pm$0.6}\\
    & Prune & 62.5    & \hspace{3pt}0.1$\pm$0.1 | 64.5$\pm$0.4 & \hspace{3pt}0.1$\pm$0.1 | 64.5$\pm$0.5 & \hspace{4pt}0.1$\pm$0.1 | 64.8$\pm$0.4 & \textbf{89.7$\pm$0.6} | \textbf{62.6$\pm$0.4}\\
    & Prune+LD & 62.5 & \hspace{3pt}0.3$\pm$0.4 | 63.7$\pm$0.5 & \hspace{3pt}0.1$\pm$0.1 | 63.8$\pm$0.5 & \hspace{4pt}0.1$\pm$0.1 | 64.1$\pm$0.3 & \textbf{84.6$\pm$0.5} | \textbf{62.8$\pm$0.2}\\
    \bottomrule
    \end{tabularx}
    \vskip -1em
    \label{tab:RQ1_table_Appendix2}
\end{table*}

\begin{table*}[h!]
    \centering
    \scriptsize
    \caption{ Results of backdooring GAT (ASR (\%) | Clean Accuracy (\%)) . Only clean accuracy is reported for clean graph.}
    \vskip -1em
    \begin{tabularx}{0.85\linewidth}{p{0.08\linewidth}p{0.04\linewidth}CCCCC}
    \toprule
    Datasets & Defense & Clean Graph & SBA-Samp & SBA-Gen & GTA & Ours  \\
    \midrule
    \multirow{3}{*}{Cora}
    & None & 84.5 & 36.8$\pm$8.7 | 85.4$\pm$1.3 & \hspace{-4pt}47.1$\pm$18.0 | 84.4$\pm$1.1 & \hspace{-4pt}72.3$\pm$27.7 | 84.6$\pm$0.8 & \textbf{99.3$\pm$0.7} | \textbf{85.4$\pm$1.0}\\
    & Prune & 81.2 & 15.7$\pm$2.4 | 83.7$\pm$0.9 & 18.9$\pm$3.6 | 83.8$\pm$0.6 & 16.6$\pm$1.5 | 83.7$\pm$1.3 & \textbf{99.6$\pm$0.1} | \textbf{83.9$\pm$1.5}\\
    & Prune+LD & 81.2 & 14.6$\pm$2.9 | 81.1$\pm$1.2 & 16.7$\pm$3.8 | 81.6$\pm$0.8 & 16.9$\pm$4.4 | 81.8$\pm$1.2 & \hspace{-4pt}\textbf{92.0$\pm$14.7} | \textbf{80.4$\pm$0.8}\\
    \midrule
    \multirow{3}{*}{Pubmed}
    & None     & 83.9 & 26.9$\pm$4.5 | 84.1$\pm$0.3 & 27.1$\pm$3.8 | 83.9$\pm$0.2 & 91.2$\pm$1.5 | 84.0$\pm$0.2 & \hspace{2pt}\textbf{100$\pm$0.0} | \textbf{84.0$\pm$4.2}\\
    & Prune    & 84.0 & 21.1$\pm$0.9 | 83.6$\pm$0.2 & 20.8$\pm$1.4 | 83.6$\pm$0.2 & 28.1$\pm$1.1 | 83.5$\pm$0.1 &\textbf{94.6$\pm$2.6} | \textbf{83.5$\pm$0.2} \\
    & Prune+LD & 84.0 & 21.5$\pm$1.3 | 83.5$\pm$0.3 & 22.0$\pm$0.8 | 83.2$\pm$0.3 & 21.8$\pm$0.4 | 83.3$\pm$0.5 &\textbf{95.3$\pm$4.1} | \textbf{84.0$\pm$4.2} \\
    \midrule
    \multirow{3}{*}{Flickr}
    & None & 46.5     & \hspace{9pt}0$\pm$0.0 | 46.9$\pm$0.2 & \hspace{9pt}0$\pm$0.0 | 46.4$\pm$0.4 & \hspace{-4pt}66.2$\pm$34.9 | 44.0$\pm$0.6 & \textbf{96.5$\pm$4.4} | \textbf{45.8$\pm$1.3} \\
    & Prune & 41.7    & \hspace{9pt}0$\pm$0.0 | 41.2$\pm$0.8 & \hspace{9pt}0$\pm$0.0 | 41.2$\pm$1.2 & \hspace{9pt}0$\pm$0.0 | 40.5$\pm$0.1 & \hspace{-4pt}\textbf{72.2$\pm$27.5} | \textbf{41.2$\pm$1.3} \\
    & Prune+LD & 41.7 & \hspace{9pt}0$\pm$0.0 | 46.1$\pm$0.6 & \hspace{9pt}0$\pm$0.0 | 46.0$\pm$0.6 & \hspace{9pt}0$\pm$0.0 | 45.7$\pm$0.4 & \textbf{96.7$\pm$6.3} | \textbf{46.0$\pm$0.3} \\
    \midrule
    \multirow{3}{*}{OGB-arxiv}
    & None & 65.3     & \hspace{3pt}1.1$\pm$1.3 | 65.4$\pm$0.1 & \hspace{-4pt}25.8$\pm$20.3 | 65.2$\pm$0.6 & 94.3$\pm$2.5 | 65.0$\pm$0.1 & \hspace{2pt}\textbf{100$\pm$0.0} | \textbf{64.8$\pm$0.2}\\
    & Prune & 61.9    & \hspace{3pt}0.1$\pm$0.1 | 63.2$\pm$0.1 & \hspace{3pt}0.1$\pm$0.1 | 63.2$\pm$0.1 & \hspace{4pt}0.1$\pm$0.1 | 63.1$\pm$0.1 & \textbf{95.5$\pm$0.0} | \textbf{62.9$\pm$0.2}\\
    & Prune+LD & 61.9 & \hspace{3pt}0.1$\pm$0.1 | 64.1$\pm$0.1 & \hspace{3pt}0.1$\pm$0.1 | 64.0$\pm$0.1 &\hspace{4pt}0.1$\pm$0.1 | 64.1$\pm$0.1 & \textbf{94.8$\pm$0.0} | \textbf{63.7$\pm$0.1}\\
    \bottomrule
    \end{tabularx}
    \vskip -1em
    \label{tab:RQ1_table_Appendix3}
\end{table*}
\begin{table}[h!]
    \centering
    \caption{Training time}
    \small
    \vskip -1em 
    \begin{tabularx}{0.7\linewidth}{XCCC}
    \toprule 
    Dataset & Size & GTA & UBGA \\
    \midrule
    Flickr & 89,250 & 18.1s & 18.3s\\ 
    Arxiv & 169,343 & 37.7s & 41.8s\\
    \bottomrule
    \end{tabularx}
    \vskip -1em
    \label{tab:time}
\end{table}
\section{Time Complexity Analysis} \label{app:complexity}
In the poisoning phase, the time complexity mainly comes from the poisoned node selection and the optimization of trigger generator. Let $h$ denote the embedding dimension. The cost of poisoned node selection with clustering is approximately $O(Kh|\mathcal{V}|)$, where $K$ is the number of clusters set in poisoned node selection and $|\mathcal{V}|$ is the number of nodes in the training graph. During the bi-level optimization phase, the computation cost of each outter iteration consists of updating of surrogate GCN model in inner iterations and training adaptive trigger generator. The cost for updating surrogate model is approximate $O(Nhd|\mathcal{V}|)$, where $d$ is the average degree of nodes and $N$ is the number of inner iterations which is generally small. The cost for optimizing the trigger generator in each outter iteration is $O(hd|\mathcal{V}|)$. Hence, the overall time complexity in each iteration of optimization is $O((N+1)hd|\mathcal{V}|)$, which is linear to the graph size. Furthermore, the framework can be trained in a mini-batch way to further reduce the computation cost in each iteration. 
In the test phase, the cost of generating trigger to attack the target node is only $O(h)$. Our time complexity analysis proves that {\method} has great potential in conducting scalable target attacks.

We also report the overall training time of our UGBA and GTA in Tab.~\ref{tab:time}. All models are trained with 200 epochs on an A6000 GPU with 48G memory.  The training time is very short and increases linearly as the complexity analysis suggests. In the test phase, attacking each target node requires 0.0017 seconds on average.

\end{document}